\documentstyle[epsfig,pra,aps]{revtex}

\def \lsim {\,{\scriptscriptstyle{\stackrel{<}{\sim}}}\,}
\def \rsim {\,{\scriptscriptstyle{\stackrel{>}{\sim}}}\,}

\newcommand{\Ref}[1]{(\ref{#1})}
\newcommand{\be}{\begin{equation}}\newcommand{\ee}{\end{equation}}
\newcommand{\bea}{\begin{eqnarray}}\newcommand{\eea}{\end{eqnarray}}
\newcommand{\beao}{\begin{eqnarray*}}\newcommand{\eeao}{\end{eqnarray*}}
\newcommand{\m}{{\,\mbox{\rm m}}}
\newcommand{\mum}{{\,\mu{\mbox{\rm m}}}}
\newcommand{\cm}{{\,\mbox{\rm cm}}}
\begin{document}

\draft
\thispagestyle{empty}
\title{
Constraints for hypothetical interactions from \\ a recent
demonstration of  the
Casimir force \\ and some possible improvements 
}

\author{
M. Bordag\footnote{Electronic address: Michael.Bordag@itp.uni-leipzig.de},
B. Geyer\footnote[5]{Electronic address: geyer@rz.uni-leipzig.de},
G.L. Klimchitskaya\footnote{on leave from North-West Polytechnical Institute,
\\St.Petersburg, Russia.  
Electronic address: \\  galina@GK1372.spb.edu}{\ }\footnotemark[4],
V.M. Mostepanenko\footnote{on leave from A.Friedmann Laboratory
for Theoretical \\ Physics, St.Petersburg, Russia. 
Electronic  address:\\ mostep@fisica.ufpb.br
}{\ }\footnote[4]{Present address: Department of Physics, 
Federal\\ University
of Paraiba, C.P. 5008, CEP 58059-970,\\ Joao Pessoa, Pb-Brazil}
}

\address{Institute for Theoretical Physics, Leipzig
University, \\ Augustusplatz 10/11, 04109 Leipzig, Germany}

\maketitle
\begin{abstract}
  The Casimir force is calculated in the configuration of a spherical lens
and a disc of finite radius covered by $Cu$ and $Au$ thin layers which
was used in a recent experiment. The correction to the Casimir force
due to finiteness of the disc radius is shown to be negligible. Also the
corrections are discussed due to the finite conductivity, large-scale
and short-scale deviations from the perfect shape of the bounding surfaces 
and the temperature correction. They were found to be essential when
confronting the theoretical results with experimental data.  Both
Yukawa-type and power-law hypothetical forces are computed which may
act in the configuration under consideration due to the exchange of light 
and/or massless elementary particles between the atoms of the lens and the 
disc.
New constraints on the constants of these forces are determined which
follow from the fact that they were not observed within the limits of
experimental errors. For Yukawa-type forces the new constraints are up to 30 
times stronger than the best ones known up today. A possible improvement of 
experimental parameters is proposed which gives the possibility to
strengthen constraints on Yukawa-type interactions up to $10^4$ times and 
on power-law interactions up to several hundred times.   
\end{abstract}

\pacs{
14.80.-j, 04.65.+e, 11.30.Pb, 12.20.Fv
}

\twocolumn

\section{INTRODUCTION}

During the past decades the Casimir effect [1] found a large
number of applications in different branches of physics (see  monograph
\cite{2} and references therein).  Among them the applications should be
mentioned in statistical physics, in elementary particle physics (e.g. in the
bag model in QCD or in Kaluza-Klein theories) and in the cosmology of the
early Universe. From the point of view of quantum field theory the Casimir
effect is a specific type of vacuum polarization which appears  by
quantizing the theory in restricted volumes or in spaces with non-trivial 
topology. This polarization
results from a change of the spectrum of zero-point oscillations  in the
presence of nontrivial boundary conditions.  For the case of the 
electromagnetic vacuum
between two uncharged metallic boundaries, separated by a small gap $a$, the
Casimir effect leads to the appearance of an attractive force acting on them
depending on $a$ and on the fundamental constants $\hbar$ and $c$ only. Such
attractive force may be alternatively explained as a retarded van der Waals
force between the two bodies whose conducting surfaces are responsible
for the Casimir force.

The Casimir force was firstly measured by Sparnaay \cite{3} for
metallic surfaces. For dielectric bodies the corresponding forces had been
measured more frequently, see [4--7]. The relative error in the
force measurements was about 100\% in \cite{3} and in the range of $(10-20)\%$
in [4--7].

As it was shown in Refs. [8,9] the Casimir force between macroscopic
bodies is very sensitive to the presence of additional hypothetical
interactions predicted by unified gauge theories,  supersymmetry and
supergravity. According to these theories interactions of two atoms arise
due to the exchange of light or massless elementary particles between them
(for example, scalar axions, graviphotons, dilatons, arions and others). Their
effective interatomic potentials may be described by Yukawa- and power-laws.
  After the integration over the volume of two macro-bodies one
obtains more complicated laws for their interaction potentials. It was rather
unexpected that quite strong constraints for the characteristic constants of
these laws may be found from the experiments on Casimir force measurements.

The constraints under consideration may be found also from other precision
experiments, e.g., from E\"{o}tv\"{o}s-, Galileo- and Cavendish-type
experiments, from the measurements of the van der Waals forces, transition
probabilities in exotic atoms etc (for a collection of references on
long-range hypothetical forces see \cite{10}).  

According to the results of
[8,11,12] the Casimir force measurements of [4--7]
lead to the strongest constraints on the constants of Yukawa-type hypothetical
interactions with a range of action of $10^{-8}\m < \lambda < 10^{-4}\m$. For
$ \lambda < 10^{-8}\m$ the best constraints follow from the measurements of
van der Waals forces in atomic force microscopy and of transition
probabilities in exotic atoms \cite{13}. For $ \lambda > 10^{-4}\m$, as it was
shown in [11,12], they follow from Cavendish- and E\"otv\"os-type
experiments [14--17].

In \cite{9} the constraints were obtained from the Casimir effect on the
constants of power-law potentials decreasing with distance as $r^{-n}$. For
$n$ = 2, 3 and 4 they turn out to be the best ones up to 1987 (compare \cite{18}).
In \cite{19} a bit stronger constraints on the power-law interaction were
obtained from the Cavendish-type experiment of Ref. \cite{14}.

Recently, a new experiment was performed \cite{20} on the measurement of the
Casimir force between two metallized surfaces of a flat disc and a spherical
lens. The absolute error of the force measurements in \cite{20} was 
$\Delta F\approx 10^{-11} N$ for distances 
$a$ between the disc and the lens in the range
$1 \mum \le a \le 6\mum$. This was the first measurement of the Casimir
force between metallic surfaces after \cite{3}. For the distance 
$a\approx 1\mum$ 
the value of the Casimir force in the configuration under consideration is
$F_C\approx -3.1\times 10^{-10}\,$N. 
This means that the relative error of the force
measurement at $1 \mum$ in \cite{20} is about $\delta\approx 3 \%$ (note that
with increasing $a$ the value of $\delta$ increases quickly). In \cite{20} the
active surfaces of the disc and the lens were covered by thin layers of copper
and gold. The use of heavier metals for  the test bodies is preferable for
obtaining stronger constraints on hypothetical interactions. This follows
from the fact that the value of the hypothetical forces increases
proportionally to the square of the density.

The aim of the present paper is to give an accurate calculation of different
hypothetical forces which might appear in the configuration used in the
experiment \cite{20}. Also, the corrections to the Casimir force due to
distortions of the surfaces, to edge effects, finite conductivity and non-zero
temperature will be analysed. On this base new constraints on the hypothetical
interactions will be reliably calculated, which follow from the results of
\cite{20}, as well as their possible improvement. 
The corresponding constraints which result
for the masses of light elementary particles are also discussed.
Some preliminary results of
this kind were obtained in \cite{21} for Yukawa-type interactions and in
\cite{22} for power-law forces. 
However, in [21,22] the corrections to the Casimir force were not
discussed and different possibilities suggested by experiment [20] were
not accounted for in full detail.

The paper is organized as follows. In Sec.~II we discuss the expression for
the Casimir force in a configuration of a plane disc and a lens and different
corrections to it are considered taking into account the finiteness of the diameter of the
disc. Sec.~III is devoted to the calculation of Yukawa-type forces in this
configuration. In Sec.~IV analogous results are obtained for the case
of power-law hypothetical forces.  Sec.~V contains a careful derivation of
the new constraints on the constants of Yukawa- and power-law interactions
which follow from the experiment performed in \cite{20}. The possible
improvement of the experiment \cite{20} is considered in Sec.~VI. For 
Yukawa-type interactions it gives the possibility to 
make the constraints about
several thousand times stronger. This considerable strengthening of
constraints may be achieved also for the power-law hypothetical interactions.
Sec.~VII contains the conclusions and some discussions. In the Appendix the
reader will find a number of mathematical details concerning the calculation
of the Casimir and hypothetical forces.

\section{THE CASIMIR FORCE BETWEEN \protect\\
A DISC AND A LENS INCLUDING \protect\\ CORRECTIONS}
The scheme of the configuration used in the recent demonstration of the
Casimir force \cite{20} is shown in Fig.~1. The Casimir force was measured
between the metallized surfaces of a flat disc (with radius
$L=1.27\cm$ and thickness  $D=0.5 \cm$) 
and a spherical lens (with
curvature radius $R=11.3 \cm$ and height $H=0.18\cm$).
The separation between them  was varied from $a=0.6\mum$ up
to $6\mum$. Both bodies were made out of quartz and covered by a continuous
layer of copper with $\Delta =0.5\mum$ thickness. The surfaces which faces
each other were covered additionally with a layer of gold of the same
thickness. Note that the penetration depth of the electromagnetic field into
gold is approximately $\delta_0\approx 0.08\mum \ll\Delta$. By this reason
when calculating the Casimir force one may consider the interacting bodies as
being made of gold as a whole.

The experimental data obtained in \cite{20} has been confronted with the
theoretical result
\begin{equation}
F_C^{(0)}(a)=-\frac{\pi^3}{360}\,R\,\frac{\hbar c}{a^3}.
\label{1}
\end{equation}
This formula is valid for the configuration of a small lens situated near the
center of a large (strictly speaking, infinite) disc at zero temperature. It
was first derived in \cite{4} and reobtained by different methods 
afterwards (see,
e.g., [2, 23]). In Ref. \cite{20}, \Ref{1} was derived from the well known
result for two infinite plane parallel plates using the Proximity Force
Theorem \cite{24}. Note, that according to our notations the attractive
forces are negative and repulsive ones are positive.

Actually, in the experiment \cite{20} the diameter of the disc was not much
larger, but even smaller than the size of the lens. Therefore it is 
of great interest
to calculate corrections to Eq. \Ref{1} due to the finite disc size.
For this purpose we use the approximation method developed earlier  for
the calculation of the Casimir force and which may be applied to the case of two
bodies with arbitrarily shaped surfaces [2, 23, 25, 26].
According to this method the potential of the Casimir force can be
obtained by summation of the retarded van der Waals interatomic potentials
over all pairs of atoms in the bodies under consideration with a subsequent
multiplicative renormalization (the latter takes into account a large amount
of the non-additivity of these forces). The method was tested and successfully
applied in the above cited papers. As a result the Casimir force may be
calculated according to:
\bea
&&F_C(a)=-\frac{\partial U_C(a)}{\partial a}, 
\label{2} \\
&&U_C(a)=-\Psi (\varepsilon_1,\varepsilon_2 )
\int\limits_{V_1}d^3 r_1
\int\limits_{V_2}d^3 r_2\, \frac{1}{r_{12}^7},
\nonumber
\eea
\noindent
where $r_{12}$ is the distance between the atoms belonging to the first and to
the second body, $\Psi (\varepsilon_1,\varepsilon_2 )$ is a tabulated
function depending on the static dielectric permittivities of the test bodies
(for its explicit form see [2, 26, 27]). In the limit of
perfectly conducting surfaces (which is of interest here)
$\varepsilon_1,\varepsilon_2\to\infty$ and $\Psi (\varepsilon_1,\varepsilon_2)
\to \pi\hbar c/24$.  In \cite{26} the relative error of the values given by
\Ref{2} was examined. It was shown to be of the order of $0,01\%$ for
configurations which do not much differ from that of two plane parallel
plates. This is just the case in the experiment \cite{20}, because only the
top of the lens and its vicinity contribute essentially to the Casimir force.

The integration in \Ref{2} for the configuration of a lens and a disc 
(Fig.~1) may be performed analytically. For this reason we introduce 
a spherical
system of coordinates with the origin in the curvature center of the lens.
The angle $\theta$ is counted from the horizontal axis directed out of the
origin to the left (see Fig.~1). Then Eq. \Ref{2} for $U_C$ takes the form
\begin{eqnarray}
&&U_C(a)=-\Psi (\varepsilon_1,\varepsilon_2 )
\int\limits_{0}^{2\pi}\!d\varphi_1\!
\int\limits_{0}^{\theta_1}\!\sin\vartheta_1\, d\vartheta_1\!\!\!
\int\limits_{R_{\min}^{\prime}(\vartheta_1)}^{R}\!\!\!r_1^2\, dr_1
\nonumber \\
&&\phantom{aaaaaa}\times
\int\limits_{0}^{2\pi}\!d\varphi_2\!
\int\limits_{0}^{\theta_2}\!\sin\vartheta_2\, d\vartheta_2\!\!\!
\int\limits_{R_{\min}(\vartheta_2)}^{R_{\max}(\vartheta_2)}\!\!\!
r_2^2\, dr_2\,
\frac{1}{r_{12}^7},
\label{3}
\end{eqnarray}
\noindent
where the integration limits are defined as follows (prime is used for 
the lens)
\begin{eqnarray}
&&R_{\min}^{\prime}(\vartheta_1)=\frac{R-H}{\cos\vartheta_1}, \quad
\theta_1=\arccos\frac{R-H}{R},
\label{4} \\
&&R_{\min}(\vartheta_2)=\frac{R+a}{\cos\vartheta_2}, \quad
R_{\max}(\vartheta_2)=\frac{R+a+D}{\cos\vartheta_2}, 
\nonumber \\
&&
\theta_2=\arccos\frac{R+a}{\sqrt{(R+a)^2+L^2}}\approx
\arccos\frac{R}{\sqrt{R^2+L^2}}.
\nonumber
\end{eqnarray}

Using the potential energy in the form of \Ref{3} with the integration limits
\Ref{4} we slightly change the experimental configuration converting the disc
into a part of a truncated cone (the corresponding addition to the disc volume
is $\Delta V/V\approx D/R\approx 4\%$). This increasing takes place, however,
near the outer boundary of the disc and practically does not influence the
result for the Casimir force.

The integrals in  (3) may be calculated along the lines presented in Appendix.
Putting $k=3$ in (A11) we obtain the result for the power-law 
interaction with a power equal to $2k+1=7$:
\bea
&&U_C(a)=-\Psi (\varepsilon_1,\varepsilon_2 )
\frac{16\pi^2}{5}
\int\limits_{0}^{\theta}\!\sin\vartheta\, d\vartheta
\label{5}\\
&&\phantom{aaaaaaaa}\times
\int\limits_{R_{\min}^{\prime}(\vartheta)}^{R}\!\!\!\!dr_1\!\!\!\!
\int\limits_{R_{\min}(\vartheta)}^{R_{\max}(\vartheta)}\!\!\!\!dr_2\,
\frac{r_1^2\, r_2^2}{(r_1+r_2)^2 (r_2-r_1)^5},
\nonumber
\eea
\noindent
where $\theta=\min(\theta_1,\theta_2)$.

We rewrite (5) in a more convenient form by introducing the new
variables $t=\cos\vartheta$, $x_1=tr_1$, $x_2=tr_2$ and using Eq.~(A6)
for $r_1 r_2$:
\begin{equation}
U_C(a)=-\Psi (\varepsilon_1,\varepsilon_2 )
\frac{\pi^2}{5}
\int\limits_{t_0}^{1}\! tdt\! 
\int\limits_{R-H}^{Rt}\!\! dx_1\!\!
\int\limits_{R+a}^{R+a+D}\!\!dx_2\,
\frac{(x_1 +x_2)^2}{(x_2-x_1)^5},
\label{6}
\end{equation}
\noindent
where $t_0=\max(R/\sqrt{R^2+L^2},(R-H)/R)$. It is seen that the value of
$t_0$ depends on the relative sizes of the lens and of the disc. If
$L\leq \sqrt{2RH}$ one has $t_0\approx 1-L^2/(2R^2)$. But if $L>\sqrt{2RH}$,
then $t_0=(R-H)/R$ and the Casimir force does not depend on a further
increase of $L$ due to the quick decreasing of the retarded van der
Waals potential with distance.

Calculating the Casimir force by the first equation of (2) the 
integration with respect to $x_2$ is removed resulting in the expression
\begin{eqnarray}
&&F_C(a)=-\Psi (\varepsilon_1,\varepsilon_2 )
\frac{\pi^2}{5}
\int\limits_{t_0}^{1}\! t\,dt\!
\int\limits_{R-H}^{Rt} \!\!dx_1
\left[
\frac{(x_1 +R+a)^2}{(R+a-x_1)^5} \right.
\nonumber \\
&&\phantom{aaaaaaa}-
\left.
\frac{(x_1 +R+a+D)^2}{(R+a+D-x_1)^5} \right].
\label{7}
\end{eqnarray}

This result coincides with (1) if we consider the limit 
$L,\varepsilon_{1,2}\to\infty$ (infinite disc made of a perfect metal) 
keeping the lowest order in the  small parameter $a/R$. Integrating eq. (7)
explicitly we find that the main contribution to the result, depending
on the size of the disc, appears at the third order in $a/R$ [28]
\begin{equation}
F_C(a) \approx
 F_C^{(0)}(a)\left[1-\frac{a^3}{R^3}\frac{1}{(1-t_0)^3}\right].
\label{8}
\end{equation}

For the parameters of experiment [20] the inequality 
$L\leq\sqrt{2RH}$ holds. In this case it follows from (8)
\begin{equation}
F_C(a) \approx F_C^{(0)}(a)\left(1-8\frac{a^3 R^3}{L^6}\right).
\label{9}
\end{equation}

It is easily seen from \Ref{9} that the correction to the Casimir force due to
the finite size of the disc does not exceed its maximal value 
$6\times 10^{-7}$ which
is achieved for $a=6\mum$. That is why one actually may neglect this
correction when confronting the measurements of the Casimir force with the
theory.

Let us discuss the corrections to the Casimir force \Ref{1} which are
significant for a small spatial separation of the lens and the disc, 
$a\approx 1\mum$. 
It is reasonable to start with the corrections due to the finite
conductivity of the metal covering the test bodies [2,29--31].
 It is well known that for $a$ in the micrometer range the
penetration depth of the electromagnetic field into the metal is frequency
independent and inversely proportional to the effective plasma frequency of the
electrons: $\delta_0=c/\omega_p$. For two plane parallel metallic
plates the corrections for the Casimir force due to the finite conductivity
were found in [29--31] (see also \cite{2}). Up to the first two
orders in the relative penetration depth $\delta_0/a$ the result is:
\begin{equation}
F_C(a) \approx
 F_C^{(0)}(a)\left(1-\frac{16}{3}\,\frac{\delta_0}{a}+
24\,\frac{\delta_0^2}{a^2}\right).
\label{10}
\end{equation}

Using the Force Proximity Theorem [24] it is not difficult to modify
(10) for the configuration of a lens and a disc:
\bea
&&F_C(a)\equiv F_C^{(0)}(a)+\Delta_{\delta_0}F_C^{(0)}(a)
\label{11} \\
&&\phantom{aaaaaa}
\approx F_C^{(0)}(a)\left(1-4\,\frac{\delta_0}{a}+
\frac{72}{5}\,\frac{\delta_0^2}{a^2}\right).
\nonumber
\eea
\noindent
Note that the first order correction in \Ref{11} was found firstly in
\cite{20}. The plus sign in front of it in \cite{20} is a misprint 
(this is also
clear from general considerations according to which $F_C(a)$ is constant 
in sign for all
$\delta_0$ and tends to zero in the formal limit $\delta_0\to\infty$ so that
the correction should be negative \cite{2}).

For gold, as it was mentioned above, $\delta_0\approx 0.08\mum$ and for
$a=1\mum$ the correction to the Casimir force $F_C^{(0)}$ due to the finite
conductivity achieves $23\%$ of $F_C^{(0)}$. The behaviour of this correction
(in relative units) with increasing $a$ is shown in Fig.~2 (curve 1). The
experimental data in \cite{20} do not support the presence of 
corrections of the result \Ref{1} being so large (let us remind that the 
relative error of the
force measurements at $1\mum$ was about $3\%$ and with such an accuracy
Eq.~(\ref{1}) was confirmed).

According to \cite{20} the reason of this contradiction is the inapplicability
of \Ref{10}, \Ref{11} for gold (in more detail, the approximation for the
effective dielectric permittivity which was used in \cite{30} to derive the
first order correction in \Ref{10} does not take into account the large
imaginary part of the refraction index for gold \cite{20}). On the other
hand the corrections due to the finite conductivity (which are in agreement
with \cite{30}) were found in \cite{31} (see also \cite{2}) in a more general
impedance approach without use of the dielectric permittivity. Consequently,
Eqs. \Ref{10} and \Ref{11} are still valid for gold covered surfaces and the
contradiction with the experimental data is still present. The most
reasonable way to resolve this problem is to take into account the
corrections of the Casimir force due to the deviations of the surfaces from the
perfect shape (in \cite{20} such corrections were not discussed).

Let us now assume that the surfaces of the lens and of the disc are covered by
some distortions with characteristic amplitudes $A_1$ and $A_2$,
correspondingly. Then the general result for the Casimir force up to second
order in the relative amplitudes of the distortions takes the form

\bea
&&F_C(a) \approx F_C^{(0)}(a)\left[1+C_{10}\frac{A_1}{a}
+C_{01}\frac{A_2}{a}\right.
\label{12} \\
&&\phantom{aaaaaaaa}\left.
+C_{20}\left(\frac{A_1}{a}\right)^2
+C_{02}\left(\frac{A_2}{a}\right)^2 +C_{11}\frac{A_1 A_2}{a^2}
\right],
\nonumber
\eea
\noindent
where the explicit expressions for the coefficients in terms of the
functions describing the shape of distortions were found in [23].

For small stochastic distortions, instead of (12), one has [23]
\bea
&&F_C(a)\equiv F_C^{(0)}(a)+\Delta_{d}F_C^{(0)}(a) 
\label{13} \\
&&\phantom{aaaaaa}
\approx F_C^{(0)}(a)\left[1+6\left(\frac{\delta_1}{a}\right)^2
+6\left(\frac{\delta_2}{a}\right)^2
\right],
\nonumber
\eea
\noindent
where $\delta_{1,2}$ are the dispersions of the stochastic perturbations on
the surfaces.  The same result is valid for the non-stochastic short-scale
distortions regardless of the specific shape of the functions describing them.
Here $\delta_{1,2}$ should be substituted by $A_{1,2}/\sqrt{2}$. Note that the
result \Ref{13} can be obtained using the Proximity Force Theorem \cite{24}
from the corresponding result for two plane parallel plates \cite{32}.

Speculating that the characteristic values of the dispersion is of order
$\delta_i\approx 0.1\mum$ we get from \Ref{13} a positive correction to the
Casimir force which is equal to $12\%$ of $F_C^{(0)}(a)$ for $a=1\mum$. In
Fig.~2 (curve 2) the dependence of this correction on $a$ is shown in
relative units.

Note that the first order correction resulting from \Ref{12} appears only in
the case when there are some large-scale deviations of boundary surfaces from
the perfect shape for which the non-perturbed force is calculated. Using a
realistic estimation of $A_i/a\approx 0.1$ for $a\approx 1\mum$ the first
order correction in \Ref{12} may amount as much as $30\%$ of $F_C^{(0)}$
\cite{23}. In the experiment \cite{20} the radius of the lens
$R=(11.3\pm0.1)\cm$ was determined with a rather large absolute error.
Consequently, a large-scale deviation of the surface of the lens from the
perfect shape could have taken place. A special inspection of the lens used
in \cite{20}, which was not done, is required to determine the actual size (and
shape) of the large-scale deviations. After that it would be possible to
calculate the coefficients $C_{10}$ and $C_{01}$ and the corresponding
correction to the Casimir force.

Together with the correction $\Delta_d F_C^{(0)}(a)$ due to the short-scale
distortions it may easily compensate the negative correction to the Casimir
force due to the finite conductivity of gold. This is possibly the reason why
in \cite{20} neither corrections to the finite conductivity nor to the 
surface distortions were observed at $a\approx 1\mum$.

One more correction to the Casimir force \Ref{1} is due to the non-zero
temperature $T$. It is calculated, e.g., in \cite{33} for two plane parallel
metallic plates (see also \cite{2}). For a lens and a disc the corresponding
result may be obtained by the use of the Proximity Force Theorem and has the
form
\bea
&&F_C(a)\equiv F_C^{(0)}(a)+\Delta_{T}F_C^{(0)}(a) 
\label{14}\\
\phantom{aaaaaa}
&&= F_C^{(0)}(a)\left[1+\frac{720}{\pi^2}f(\xi)
\right].
\nonumber
\eea
\noindent
Here $\xi=k_B Ta/(\hbar c)$, $k_B$ is Boltzmann's constant and
\begin{eqnarray}
&&f(\xi)=\frac{1}{4\pi^2}\sum\limits_{l,m=1}^{\infty}
\frac{(2\xi)^4}{\left[l^2+m^2(2\xi)^2\right]^2}
\label{15} \\
&&\phantom{a}
=\frac{1}{4\pi^2}\left\{\sum\limits_{l=1}^{\infty}
\left[\frac{2\pi\xi^3}{l^3}
\frac{\cosh\frac{\pi l}{2\xi}}{\sinh\frac{\pi l}{2\xi}}+
\frac{(\pi\xi)^2}{l^2}\sinh^{-2}\left(\frac{\pi l}{2\xi}\right)\right]
\right.
\nonumber \\
&&\phantom{aaaaaaaaaaaa}
\left.-
\frac{4\pi^4}{45}\xi^4
\vphantom{\left[
\frac{\cosh\frac{\pi l}{2\xi}}{\sinh\frac{\pi l}{2\xi}}\right]}
\right\}.
\nonumber
\end{eqnarray}

The relative value of the temperature corrections, calculated with \Ref{14},
\Ref{15} is shown in Fig.~2 (curve 3) in dependence on $a$. It is seen that
for $a=1\mum$ it is approximately $2.7\%$ of $F_C^{(0)}$. But for 
$a=6\mum$
the temperature correction is $\Delta_T F_C^{(0)}=1.74 F_C^{(0)}$.

Fig.~2 will be used in Secs.~V,~VI which are devoted to obtain stronger
constraints on the constants of hypothetical interactions from the Casimir
force measurements (the other forces which may contribute in the experiment,
i.e. electric one
should be subtracted to get the result for the Casimir force with possible
corrections discussed above, see \cite{20}).

\section{THE YUKAWA-TYPE HYPOTHETICAL \protect\\ INTERACTION 
BETWEEN A DISC \protect\\ AND A LENS}

The Yukawa potential between two atoms of the interacting bodies may be
represented in the form
\begin{equation}
V_{Yu}=-\alpha\, N_1\, N_2\,\hbar\, c\,\frac{1}{r_{12}}
{e}^{-r_{12}/\lambda},
\label{16}
\end{equation}
\noindent
where $\alpha$ is a dimensionless interaction constant,
$\lambda=\hbar /(mc)$ is the Compton wavelength of a hypothetical
particle which is responsible for the rise of new interactions,
and $N_i$ are the numbers of nucleons in the atomic nuclei;
they are introduced in (16) to make $\alpha$ independent on the sort
of atoms.

The potential energy of hypothetical interaction in the configuration of
experiment [20] (see Fig.~1) may be obtained as the additive sum of
the potentials (16) with appropriate atomic densities of the lens and
the disc materials
\begin{eqnarray}
&&U_{Yu}(a)=-\alpha\,\frac{\hbar c}{m_p^2}
\sum\limits_{i,j=1}^{3}\rho_i^{\prime}\rho_j U_{Yu}^{(i,j)}(a),
\nonumber \\
&& U_{Yu}^{(i,j)}(a)=
\int\limits_{V_i^{\prime}}d^3 r_1
\int\limits_{V_j}d^3 r_2\,
\frac{1}{r_{12}}\,{e}^{-r_{12/\lambda}}.
\label{17}
\end{eqnarray}
\noindent
Here $\rho_i^{\prime}$, $V_i^{\prime}$ ($i=1,2,3$)
 are the densities and volumes of the lens and the covering metallic
layers ($\rho_j$, $V_j$ are the same for the disc). The proton mass
$m_p$ appeared due to the use of usual densities instead of the atomic
ones. In numerical calculations of Sec.~V the values
$\rho_1^{\prime}=2.23\,$g/cm$^3$,
$\rho_1=2.4\,$g/cm$^3$,
$\rho_2^{\prime}=\rho_2=8.96\,$g/cm$^3$,
$\rho_3^{\prime}=\rho_3=19.32\,$g/cm$^3$ [20] will be used.  

The hypothetical force between lens and disc can be computed
as the derivative
\begin{equation}
F_{Yu}(a)=-\frac{\partial U_{Yu}(a)}{\partial a}.
\label{18}
\end{equation}

The integration (17) can be performed most simply in a spherical coordinate
system described in Sec.~II (see Fig.~1). In these coordinates the multiple
integral from (17) takes the form

\begin{eqnarray}
&&U_{Yu}^{(i,j)}(a)=
\int\limits_{0}^{2\pi}\!d\varphi_1\!
\int\limits_{0}^{\theta_1}\!\sin\vartheta_1\, d\vartheta_1\!\!
\int\limits_{R_{i,\min}^{\prime}}^{R_{i,\max}^{\prime}}
\!\!\!r_1^2\, dr_1
\label{19} \\
&&\phantom{aaaaaa}\times
\int\limits_{0}^{2\pi}\!d\varphi_2\!
\int\limits_{0}^{\theta_2}\!\sin\vartheta_2\, d\vartheta_2\!\!
\int\limits_{R_{j,\min}}^{R_{j,\max}}\!\!\!r_2^2\, dr_2\,
\frac{1}{r_{12}}\,
{e}^{-r_{12}/\lambda}.
\nonumber
\end{eqnarray}
\noindent
The quantities $\theta_1$, $\theta_2$ were defined in (4), the other
integration limits are (as above by the prime the lens parameters
are notated)
\begin{eqnarray}
&&
R_{1,\min}^{\prime}=\frac{R-H}{\cos\vartheta_1},\quad
R_{1,\max}^{\prime}=R_{2,\min}^{\prime}=
R-\Delta_1^{\prime}-\Delta_2^{\prime},
\nonumber \\
&&R_{2,\max}^{\prime}=R_{3,\min}^{\prime}=R-\Delta_2^{\prime},
\qquad R_{3,\max}^{\prime}=R,
\label{20} \\
&&R_{1,\min}=R_{2,\max}=
\frac{R+a+\Delta_1+\Delta_2}{\cos\vartheta_2}, 
\nonumber \\
&&
R_{1,\max}=\frac{R+a+D}{\cos\vartheta_2},\qquad
R_{3,\min}=\frac{R+a}{\cos\vartheta_2},
\nonumber \\
&&R_{2,\min}=R_{3,\max}=
\frac{R+a+\Delta_2}{\cos\vartheta_2}. 
\nonumber
\end{eqnarray}
\noindent
For generality the thicknesses of metallic layers on the lens and on the disc
are permitted to be different.

To calculate the integrals in (19) it is convenient to use the expansion
into a series of spherical  harmonics [34]
\bea
&&\frac{{e}^{-r_{12}/\lambda}}{r_{12}}=\frac{4\pi}{\sqrt{r_1 r_2}}
\sum\limits_{l=0}^{\infty}
{I}_{l+\frac{1}{2}}\left(\frac{r_1}{\lambda}\right)\,
{K}_{l+\frac{1}{2}}\left(\frac{r_2}{\lambda}\right)\,
\label{21}\\
&&\phantom{aaaaaa}\times
\sum\limits_{m=-l}^{l}
{Y}_{lm}^{\ast}(\vartheta_1,\varphi_1)\,
{Y}_{lm}(\vartheta_2,\varphi_2),
\nonumber
\eea
\noindent
where ${I}_{\nu}(z)$, ${K}_{\nu}(z)$ are Bessel functions of
imaginary argument. For large arguments ($z\gg 1$) one has  
\begin{equation}
{I}_{l+\frac{1}{2}}(z)\approx\frac{1}{\sqrt{2\pi z}}\,{e}^z, 
\qquad
{K}_{l+\frac{1}{2}}(z)\approx\sqrt{\frac{\pi}{2z}}\,{e}^{-z}.
\label{22}
\end{equation}

Let us consider separately the cases of small and large parameter
$\lambda$. In (21) the condition $z\gg 1$ corresponds to
$\lambda\ll r_1, r_2$ which is valid when  $\lambda\ll R$
(actually $\lambda$ must be less than the lowest size of the interacting 
bodies, i.e. $\lambda < H$ in the configuration under consideration).
Substituting the asymptotics (22) into the right-hand side of (21) and
taking into account the completeness relation for spherical harmonics
(A9) we obtain for $\lambda < H$:
\bea
&&\frac{{e}^{-r_{12}/\lambda}}{r_{12}}\approx
\frac{2\pi\lambda}{r_1\, r_2}\,{e}^{r_1/\lambda}\,{e}^{-r_2/\lambda}
\label{23} \\
&&\phantom{aaaaaaa}\times
\delta(\varphi_1-\varphi_2)\,\delta(\cos\vartheta_1-\cos\vartheta_2).
\nonumber
\eea

Substituting (23) into (19) we get two different situations depending 
on the value of index $i$. If $i=2,3$ (the integration is over the layers 
covering the lens) the integration limits in $r_1$ do not depend on
$\vartheta$: 
\bea
&&U_{Yu}^{(i,j)}(a)=4\pi^2\lambda
\int\limits_{0}^{\theta}\!\sin\vartheta\, d\vartheta\!\!
\label{24} \\
&&\phantom{aaaaaaaaa}\times
\int\limits_{R_{i,\min}^{\prime}}^{R_{i,\max}^{\prime}}\!\!\! 
r_1\,dr_1\,{e}^{r_1/\lambda}\!\!\!\! 
\int\limits_{R_{j,\min}(\vartheta)}^{R_{j,\max}(\vartheta)}\!\!\!\! 
r_2\,dr_2\,{e}^{-r_2/\lambda},
\nonumber
\eea
\noindent
where $\theta=\min(\theta_1,\theta_2)$.

It is convenient to consider the force
$f_{Yu}^{(i,j)}(a)=-\partial U_{Yu}^{(i,j)}(a)/\partial a$ instead of the
potential energy (24). This gives the possibility to remove the
integration with respect to $r_2$. Integrating with respect to the variables $r_1$ and $\vartheta$
in a standard way [35] one gets for $i=2,3$ and $j=1,2,3$:
\begin{eqnarray}
&&f_{Yu}^{(i,j)}(a)=4\pi^2\lambda^3
{e}^{-a_{ij}/\lambda}(R-\lambda)
\left(1-{e}^{-\Delta_{i-1}^{\prime}/\lambda} \right) 
\label{25} \\
&&\phantom{aaaaa}\times
\left\{1-{e}^{-BR/\lambda}-
{e}^{-d_j/\lambda} \left[1-{e}^{-(R+d_j)B/\lambda}\right]\right\}.
\nonumber
\end{eqnarray}
\noindent
Here the following notations are used:
\begin{eqnarray}
&&a_{33}=a, \qquad a_{32}=a+\Delta_1, \qquad 
a_{31}=a+\Delta_1+\Delta_2, 
\nonumber \\
&&a_{23}=a+\Delta_2^{\prime},\qquad
a_{22}=a+\Delta_2+\Delta_2^{\prime},
\nonumber \\
&& 
a_{21}=a+\Delta_2+\Delta_1^{\prime}+\Delta_2^{\prime},
\label{26} \\
&&d_1=D-\Delta_1-\Delta_2\approx D, \qquad 
d_2=\Delta_1, \qquad d_3=\Delta_2, 
\nonumber \\
&&B=\min\left(\frac{L^2}{2R^2},\frac{H}{R}\right).  
\nonumber
\end{eqnarray}

Now we consider the second situation where the integration in (19) is 
over the lens itself ($i=1$). Here the quantity $U_{Yu}^{(1,j)}(a)$
has the same form as in (24) but with integration limit
$R_{1,\min}^{\prime}$ depending on $\vartheta$. Considering the force
instead of potential energy and performing the integration by the use
of tabulated formulas [35] we find the result:
\begin{eqnarray}
&&f_{Yu}^{(1,j)}(a)=4\pi^2\lambda^3
{e}^{-a_{1j}/\lambda}\left\{(R-\lambda)
\vphantom{\left[{e}^{-(H+d_j)B/\lambda}\right]}
\right.
\label{27} \\
&&\phantom{aaaa}\times
\left[1-{e}^{-BR/\lambda}
-{e}^{-d_j/\lambda} \left(1-{e}^{-(R+d_j)B/\lambda}\right)\right]
\nonumber \\
&&\phantom{aaaa}
-{e}^{-H/\lambda} \left[ A_1-A_2{e}^{-BH/\lambda}
-A_3^{(j)}{e}^{-d_j/\lambda}
\vphantom{{e}^{-(H+d_j)B/\lambda}}
\right.
\nonumber \\
&&\phantom{aaaaaaaaa}\left.\left.
+
A_4^{(j)}{e}^{-d_j/\lambda}{e}^{-(H+d_j)B/\lambda}\right]\right\}.
\nonumber
\end{eqnarray}
\noindent
Here the following notations are introduced:
\begin{eqnarray}
&&a_{11}=a+\Delta_1+\Delta_2+\Delta_1^{\prime}+\Delta_2^{\prime}, 
\nonumber \\ 
&&a_{12}=a+\Delta_2+\Delta_1^{\prime}+\Delta_2^{\prime},  
\nonumber \\ 
&&a_{13}=a+\Delta_1^{\prime}+\Delta_2^{\prime}, 
\label{28} \\ 
&&A_1=\frac{R}{H}\left[\frac{(R-H)(H+\lambda)}{H}-\lambda\right], 
\nonumber \\ 
&&
A_2=\frac{R}{H}\left[\frac{(R-H)(H+\lambda t_0)}{H}-\lambda\right], 
\nonumber \\
&&A_3^{(j)}=\frac{R+d_j}{H+d_j}\left[\frac{(R-H)(H+d_j+\lambda)}{H}
-\lambda\right], 
\nonumber \\
&&A_4^{(j)}=\frac{R+d_j}{H+d_j}\left[\frac{(R-H)(H+d_j+\lambda t_0)}{H}
-\lambda\right],
\nonumber
\end{eqnarray}
\noindent
$t_0$ is defined in explanations to (6).

The complete value of the hypothetical force, according to (17), (18),
may be presented in the form:
\begin{equation}
F_{Yu}=-\alpha\,\frac{\hbar c}{m_p^2}
\sum\limits_{i,j=1}^{3}\rho_i^{\prime}\,\rho_j\, f_{Yu}^{(i,j)},
\label{29}
\end{equation}
\noindent
where $f_{Yu}^{(i,j)}$ are defined in (25), (27). The expressions (25), 
(27), (29) will be used in Secs.~V,VI for calculating constraints on
Yukawa-type hypothetical interactions.

Note that for the extremely small
$\lambda{\,{\scriptscriptstyle{\stackrel{<}{\sim}}}\,}a$
the expression (29) may be additionally simplified:
\begin{eqnarray}
&&F_{Yu}=-\alpha\frac{\hbar c}{m_p^2}4\pi^2\lambda^3
{e}^{-a/\lambda}R
\label{30} \\
&&\phantom{aaaa}\times
\left[\rho_1^{\prime}{e}^{-(\Delta_1^{\prime}+\Delta_2^{\prime})/\lambda}
+\rho_2^{\prime}{e}^{-\Delta_2^{\prime}/\lambda}+\rho_3^{\prime}\right]
\nonumber \\
&&\phantom{aaaaaa}\times
\left[\rho_1{e}^{-(\Delta_1+\Delta_2)/\lambda}
+\rho_2
{e}^{-\Delta_2/\lambda}+\rho_3\right].
\nonumber
\end{eqnarray}
\noindent
Exactly this result (for equal thicknesses of the covering metallic layers)
follows in the limit of small $\lambda$ from the formula derived in [21]
for the configuration of a lens and an infinite disc. Thus for
$\lambda{\,{\scriptscriptstyle{\stackrel{<}{\sim}}}\,}a$
the finiteness of the disc size does not influence the value of
the hypothetical force. At the same time for larger $\lambda$ it is
necessary to take into account the finite sizes of the disc (unlike
the case when we calculated the Casimir force in Sec.~II). 

Let us start with the case of large $\lambda$ ($\lambda\gg R$). 
Now we may neglect the exponent in Eq.~(16). As a result 
the interatomic potential does not depend on $\lambda$ any more and behaves 
as $1/r_{12}$. For this potential the hypothetical force between a lens
and a disc was calculated in [9] under the assumption that the disc area
is infinitely large. Such suggestion is not reliable for the potential
under consideration due to its slow decrease with the distance. The
general method developed in the Appendix for the potentials of the form
$r_{12}^{-(2k+1)}$ with $k\geq 1$ also should be modified to include the
case $k=0$. We will take into account that for large $\lambda$ the
covering metallic layers practically do not contribute to the result.
By this reasoning one may integrate directly over the lens and the disc 
volumes.
As a result in a spherical coordinate system used above the expression for
the hypothetical force is:
\begin{eqnarray}
&&F_{Yu}\approx -\alpha\,\frac{\hbar c}{m_p^2}\,\rho_1^{\prime}\,\rho_1
\int\limits_{0}^{2\pi}\!d\varphi_1
\int\limits_{0}^{2\pi}\!d\varphi_2
\int\limits_{0}^{\theta_1}\!\sin\vartheta_1\, d\vartheta_1
\label{31} \\
&&\phantom{aa}\times
\int\limits_{0}^{\theta_2}\!\sin\vartheta_2\, d\vartheta_2\!\!
\int\limits_{R_{\min}^{\prime}}^{R}\!\!\!r_1^2\, dr_1\!\!\!
\int\limits_{R_{\min}}^{R_{\max}}\!\!\!r_2^2\, dr_2\,
\frac{r_2\cos\vartheta_2-r_1\cos\vartheta_1}{r_{12}^3}.
\nonumber
\end{eqnarray}

The calculational details for integration in (31) are given in Appendix.
The result is:
\bea
&&F_{Yu}\approx -\alpha\frac{\hbar c}{m_p^2}\rho_1^{\prime}\rho_1
2\pi^2L^2DH
\label{32} \\
&&\phantom{aaaa}\times
\left[1-\frac{L^2}{4RH}\left(\ln\frac{D+H}{D}+
\frac{H}{D}\ln\frac{D+H}{H}\right)\right].
\nonumber
\eea

In the intermediate range between $\lambda < H$ and $\lambda\gg R$ the
integration in (17) was performed numerically. 
For this purpose algorithm 698 from netlib [36] was used. It is an adaptive
multidimensional integration, the FORTRAN program is called DCUHRE.
A large number of function calls (about $5\times 10^6$) for
$\lambda\geq 2.3\times 10^{-4}\,$m was necessary in order to obtain
reliable results. For smaller values $\lambda$ the program does not work.
The results of numerical calculations are in good agreement with the 
analytical ones (see Secs.~V,VI).


\section{CALCULATION OF A POWER-LAW \protect\\ INTERACTION}

In this Section the force is calculated which may act in the configuration
of experiment [20] due to power-law hypothetical interactions. The power-law 
potential between two atoms belonging to a lens and a disc is
\be
V_n=-\lambda_n\,N_1\,N_2\,\hbar\,c\,\frac{1}{r_{12}}\left(
\frac{r_0}{r_{12}}\right)^{n-1}.
\label{33}
\ee
\noindent
Here $\lambda_n$ is a dimensionless constant and $r_0=1$F$\,=10^{-15}\,$m
is introduced for the proper dimensionality of potentials with different 
$n$ [18].

The potential energy of the lens and the disc (see Fig.~1) due to a hypothetical
interaction is an additive sum of the potentials (33) with appropriate
atomic densities. It can be written in a form analogous to (17)
\bea
&&U_{n}(a)=-\lambda_n \frac{\hbar\, c\, r_0^{n-1}}{m_p^2}
\sum\limits_{i,j=1}^{3}\rho_i^{\prime}\rho_j U_{n}^{(i,j)}(a),
\label{34} \\
&&F_n(a)=-\frac{\partial U_n(a)}{\partial a}.
\nonumber
\eea
\noindent
The quantities $U_{n}^{(i,j)}$ here are defined by the same integrals
as in (17) where instead of $\exp(-r_{12}/\lambda)$ the function
$1/r_{12}^{n-1}$ should be substituted.

At first we consider the case $n=3$. Here the result for $U_{3}^{(i,j)}$
is given by (A11) with $k=1$. We rewrite it by using (A6) for $r_1 r_2$
and introducing the new variables $t=\cos\vartheta$, $x_1=tr_1$,
$x_2=tr_2$
\be
U_{3}^{(i,j)}(a)=\pi^2
\int\limits_{t_0}^{1}\frac{dt}{t^3} 
\int\limits_{tR_{i,\min}^{\prime}}^{tR_{i,\max}^{\prime}}\!\!dx_1\!\!
\int\limits_{tR_{j,\min}}^{tR_{j,\max}}\!\!dx_2\,
\frac{(x_1+x_2)^2}{x_2-x_1},
\label{35}
\ee
\noindent
where $t_0$ is defined in the explanation to (6).

Calculating the force, one can eliminate the integration with respect to $x_2$:
\bea
&&F_{3}^{(i,j)}(a)=-\frac{\partial U_{3}^{(i,j)}(a)}{\partial a}
\label{36} \\
&&\phantom{a}=\pi^2\!
\int\limits_{t_0}^{1}\frac{dt}{t^3}\! 
\int\limits_{tR_{i,\min}^{\prime}}^{tR_{i,\max}^{\prime}}\!\!\!\!dx_1
\left[\frac{(x_1+tR_{j,\max})^2}{tR_{j,\max}-x_1}-
\frac{(x_1+tR_{j,\min})^2}{tR_{j,\min}-x_1}\right].
\nonumber
\eea

Performing the integration in (36) for different values of $i,j$ to
lowest order in the small parameters $H/R$, $L^2/(2R^2)$, $D/R$,
$\Delta_i^{\prime}/H$, $\Delta_j/D$, $a/H$, $a/D$ and $a/L$, and
substituting the results into (34), we obtain:
\bea
&&
F_3(a)=-4\pi^2\lambda_3 \frac{\hbar\, c\, r_0^{2}}{m_p^2}\,R
\label{37} \\
&&\phantom{aa}\times
\left\{\rho_1^{\prime}\rho_1\left[D\ln\frac{2RD+\tilde{L}^2}{2RD}+
\frac{\tilde{L}^2}{2R}\ln\frac{(2RD+\tilde{L}^2)H}{\tilde{L}^2(H+D)}
\right]\right.
\nonumber \\
&&\phantom{aaaa}+\rho_1^{\prime}\left[\rho_2\Delta_1
\ln\frac{\tilde{L}^2}{2R(a+\Delta_1+\Delta_2+\Delta_1^{\prime}
+\Delta_2^{\prime})}\right.
\nonumber \\
&&\phantom{aaaaaaaaaaaaaa}
\left.+\rho_3\Delta_2
\ln\frac{\tilde{L}^2}{2R(a+\Delta_2+\Delta_1^{\prime}+
\Delta_2^{\prime})}\right]
\nonumber \\
&&\phantom{aaaa}+\rho_1\left(\rho_2^{\prime}\Delta_1^{\prime}
\ln\frac{D}{a+\Delta_1+\Delta_2+\Delta_2^{\prime}}
\right.
\nonumber \\
&&\phantom{aaaaaaaaaaaaaa}
\left.+
\rho_3^{\prime}\Delta_2^{\prime}
\ln\frac{D}{a+\Delta_1+\Delta_2}\right)
\nonumber \\
&&\phantom{aaaa}+\rho_2\left(\rho_2^{\prime}\Delta_1^{\prime}
\ln\frac{a+\Delta_1+\Delta_2+\Delta_2^{\prime}}{a+\Delta_2+\Delta_2^{\prime}}
\right.
\nonumber \\
&&\phantom{aaaaaaaaaaaaaa}
\left.
+\rho_3^{\prime}\Delta_2^{\prime}
\ln\frac{a+\Delta_1+\Delta_2}{a+\Delta_2}\right)
\nonumber \\
&&\phantom{aaaa}\left.
+\rho_3\left(\rho_2^{\prime}\Delta_1^{\prime}
\ln\frac{a+\Delta_2+\Delta_2^{\prime}}{a+\Delta_2^{\prime}}
+\rho_3^{\prime}\Delta_2^{\prime}
\ln\frac{a+\Delta_2}{a}\right)
\vphantom{\left[\frac{\tilde{L}^2}{2R}\right]}
\right\}.
\nonumber
\eea
\noindent
Here $\tilde{L}^2\equiv\min(L^2,2RH)$. Note that in the specific case
$\Delta_1=\Delta_2=\Delta_1^{\prime}=\Delta_2^{\prime}=\Delta$,
$\rho_2^{\prime}=\rho_2$, $\rho_3^{\prime}=\rho_3$,
and $\tilde{L}=L$ Eq.~(37) coincides with
Eq.~(16) of Ref.~[22].

Let us discuss now the power-law potentials with the even powers $n=2,4$.
In this case the corresponding quantities $U_{2k}^{(i,j)}$ ($k=1,2$)
from Eq.~(34) are most conveniently represented in cylindrical
coordinates ($\tau,\,\varphi,\,z$) with the origin at the lens top and
the $z$-axis orthogonal to the disc surface. Then the quantities
$U_{2k}^{(1,j)}$ describing the interaction energy of the lens atoms with the
atoms of the disc and its covering layers take the form
\bea
&&
U_{2k}^{(1,j)}(a)=4\pi\!
\int\limits_{\Delta_1^{\prime}+\Delta_2^{\prime}}^{H}\!\!\!\!dz_1\!
\int\limits_{0}^{f(z_1)}\!\!d\tau_1\!\!\!
\int\limits_{-a_j-d_j}^{-a_j}\!\!\!\!dz_2
\int\limits_{0}^{L}\!d\tau_2
\label{38} \\
&&\phantom{aaaaaaaa}\times
\int\limits_{0}^{\pi}\frac{\tau_1\,\tau_2\,d\varphi}{\left[\tau_1^2+\tau_2^2-
2\tau_1\tau_2\cos\varphi+(z_1-z_2)^2\right]^k}.
\nonumber
\eea
\noindent
In (38) the following notations are introduced
\bea
&&f(z_1)=\left[2(R-\Delta_1^{\prime}-\Delta_2^{\prime})z_1-
z_1^2\right]^{\frac{1}{2}},
\label{39} \\
&&a_1=a+\Delta_1+\Delta_2, \qquad a_2=a+\Delta_2, \qquad a_3=a,
\nonumber
\eea
\noindent
the thicknesses $d_j$ were defined in (26).

 The quantities
$U_{2k}^{(i,j)}$ with $i=2,3$ describe the interaction energy of the lens 
covering layers with the disc and its layers.
They are expressed by
\bea
&&
U_{2k}^{(1,j)}(a)=4\pi
\int\limits_{0}^{\theta_1} \sin\vartheta_1\,d\vartheta_1\!\!
\int\limits_{R_{i,\min}^{\prime}}^{R_{i,\max}^{\prime}}\!\!\!\!dr_1\!\!
\int\limits_{-a_j-d_j}^{-a_j}\!\!\!\!dz_2
\int\limits_{0}^{L}\!d\tau_2
\label{40} \\
&&\phantom{aaaaaaaa}\times
\int\limits_{0}^{\pi}\frac{r_1^2\,\tau_2\,d\varphi}{\left[\tau_1^2+\tau_2^2-
2\tau_1\tau_2\cos\varphi+(z_1-z_2)^2\right]^k}.
\nonumber
\eea
\noindent
Here the layers covering the lens are described in spherical
coordinates used above, so that $\tau_1=r_1\sin\vartheta_1$,
$z_1=R-r_1\cos\vartheta_1$.

To perform the integration with respect to $\varphi$ in (38), (40)
it is helpful to use the integral representation of Legendre
polynomials [35]. Also the integration with respect to $z_2$ can be
removed when calculating the force instead of the potential energy.
The result for $i=1$ is
\bea
&&F_{2k}^{(1,j)}(a)=
\int\limits_{\Delta_1^{\prime}+\Delta_2^{\prime}}^{H}\!\!\!dz_1\!
\int\limits_{0}^{f(z_1)}\!\!\tau_1\,d\tau_1
\int\limits_{0}^{L}\tau_2\,d\tau_2
\label{41} \\
&&\phantom{aaaaaaaaaa}\times
\sum\limits_{p=1}^{2}
\frac{(-1)^p}{(A_{1j}^{(p)})^k}\,P_{k-1}(z_{1j}^{(p)}).
\nonumber
\eea

In the same way  for $i=2,3$ the result is 
\bea
&&F_{2k}^{(i,j)}(a)=
\int\limits_{0}^{\theta_1} \!\sin\vartheta_1\,d\vartheta_1\!\!
\int\limits_{R_{i,\min}^{\prime}}^{R_{i,\max}^{\prime}}\!\!\!r_1^2\,dr_1\!
\int\limits_{0}^{L}\!\tau_2\,d\tau_2
\label{42} \\
&&\phantom{aaaaaaaaaa}\times
\sum\limits_{p=1}^{2}
\frac{(-1)^p}{(A_{ij}^{(p)})^k}\,P_{k-1}(z_{ij}^{(p)}).
\nonumber
\eea
\noindent
In (41), (42) the following notations are introduced
\bea
&&A_{1j}^{(p)}=\left\{
\left[\tau_1^2+\tau_2^2+(z_1+h_j^{(p)})^2\right]^2-4\tau_1^2\tau_2^2
\right\}^{\frac{1}{2}},
\nonumber \\
&&z_{1j}^{(p)}=
\frac{1}{A_{1j}^{(p)}}\left[\tau_1^2+\tau_2^2+(z_1+h_j^{(p)})^2\right],
\label{43} \\
&&h_j^{(1)}=a_j, \qquad h_j^{(2)}=a_j+d_j.
\nonumber 
\eea
\noindent
The quantities $A_{ij}^{(p)},z_{ij}^{(p)}$ with $i=2,3$ are obtained from
$A_{1j}^{(p)},z_{1j}^{(p)}$ by substitution of $\tau_1, z_1$ according to
explanations after Eq.~(40).

Eqs.~(41), (42) with the notations (43) look rather cumbersome. In spite of
this all involved integrals can be calculated explicitly by the use of
Ref.~[35]. As a result the hypothetical force $F_4(a)$ computed according
(34), (41), (42) in the lowest order in small parameters mentioned above is:
\bea
&&
F_4(a)=-2\pi^2\lambda_4 \frac{\hbar\, c\, r_0^{3}}{m_p^2}\,R
\label{44} \\
&&\phantom{aaa}\times
\left\{\rho_1^{\prime}\rho_1\left[
\ln\frac{{L}^2}{2R(a+\Delta_1+\Delta_2+\Delta_1^{\prime}+
\Delta_2^{\prime})}-\frac{{L}^2}{2RH}
\right]\right.
\nonumber \\
&&\phantom{aaaaa}+\rho_1^{\prime}\left(\rho_2
\ln\frac{a+\Delta_1+\Delta_2+\Delta_1^{\prime}+
\Delta_2^{\prime}}{a+\Delta_2+\Delta_1^{\prime}+
\Delta_2^{\prime}}\right.
\nonumber \\
&&\phantom{aaaaaaaaaaaa}\left.
+\rho_3\ln\frac{a+\Delta_2+\Delta_1^{\prime}+
\Delta_2^{\prime}}{a+\Delta_1^{\prime}+\Delta_2^{\prime}}
\right)
\nonumber \\
&&\phantom{aaaaa}+\rho_1\left(\rho_2^{\prime}
\frac{\Delta_1^{\prime}}{a+\Delta_1+\Delta_2+\Delta_2^{\prime}}+
\rho_3^{\prime}
\frac{\Delta_2^{\prime}}{a+\Delta_1+\Delta_2}\right)
\nonumber \\
&&\phantom{aaaaa}
+\rho_2^{\prime}
\frac{\Delta_1^{\prime}}{a+\Delta_2+\Delta_2^{\prime}}
\left(\rho_2\frac{\Delta_1}{a+\Delta_1+\Delta_2+\Delta_2^{\prime}}
\right.
\nonumber \\
&&\phantom{aaaaaaaaaaaaaaa}\left.+
\rho_3\frac{\Delta_2}{a+\Delta_2^{\prime}}\right)
\nonumber \\
&&\phantom{aaaaa}\left.
+\rho_3^{\prime}
\frac{\Delta_2^{\prime}}{a+\Delta_2}
\left(\rho_2\frac{\Delta_1}{a+\Delta_1+\Delta_2}+
\rho_3\frac{\Delta_2}{a}\right)
\vphantom{\left[\frac{{L}^2}{2R}\right]}
\right\}.
\nonumber
\eea

The hypothetical force $F_2(a)$ calculated by
(34), (41), (42) in the lowest order of the same small parameters is:
\bea
&&
F_2(a)=-2\pi^2\lambda_2 \frac{\hbar\, c\, r_0}{m_p^2}\,
\rho_1^{\prime}\rho_1\left[
\frac{D\,{L}^2}{2}-
D^2\,R\ln\frac{2R\,D+{L}^2}{2R\,D}
\right.
\nonumber \\
&&\phantom{aaaa=}-
\frac{{L}^2}{2R^2}(L^2\,R-D^2\,R-DL^2)\ln\frac{2R\,D+{L}^2}{L^2}
\nonumber  \\
&&\phantom{aaaa=}+
L^2\,H\ln\frac{D+H}{H}+L^2\,D\ln\frac{2R(D+H)}{2D\,R+L^2}
\label{45}\\
&&\phantom{aaaa=}\left.+
\frac{{L}^2\,D}{4R^2}(L^2+D\,R)\ln\frac{L^4}{2R^2(2R\,H-{L}^2)}
\right].
\nonumber
\eea

Note, that here only the contribution of the lens and the disc materials is
written out. The covering metallic layers practically do not contribute
to the value of the force for the power-law potentials with $n=2$ [22].


\section{CONSTRAINTS FOR HYPOTHETICAL \protect\\ INTERACTIONS FROM THE 
RECENT \protect\\ EXPERIMENT}

The results of Secs.~II--IV are used here for obtaining stronger constraints 
on the constants of hypothetical 
Yukawa- and power-type interactions which follow from
the measurements of experiment [20]. The absolute error of the force
 measurements in [20] was $\Delta F=1\,\mu$dyn$\,=10^{-11}\,$N in a range of distance
 between the lens and the disc from $a=1\,\mu$m till $a=6\,\mu$m. With
this error the expression (1) for the Casimir force was confirmed and
no corrections to it or unexpected interactions were observed.

We now discuss the values of $a$ for which the most strong and reliable
constraints on hypothetical interactions can be obtained from the 
above mentioned result.
Thereby the corrections to the Casimir force considered in
Sec.~II have to be taken into account. 
Evidently, 
because all 
kinds of hypothetical forces decrease with distance
the smallest values of $a$ are to be prefered. From this point of view, 
e.g., $a=1\,\mu$m should be chosen. But for these values of $a$, as it
follows from Sec.~II, the theoretical value of the force under measuring
is not strictly defined. Although the Casimir force itself is rather large
($F_C^{(0)}\approx -3.1\times 10^{-10}\,$N) and the temperature
correction to it is rather small 
($\Delta_{T}F_C^{(0)}\approx -0.8\times 10^{-11}\,$N) the corrections due
to finite conductivity and due to surface distortions are large. According
to Fig.~2 (curve~1) the corrections to finite conductivity
$\Delta_{\delta_0}F_C^{(0)}$ at $a=1\,\mu$m is of order
$-0.23F_C^{(0)}\approx 7.1\times 10^{-11}\,$N. The correction due to
short-scale distortions $\Delta_{d}F_C^{(0)}$ (curve~2) is of order
$0.12F_C^{(0)}\approx -3.7\times 10^{-11}\,$N. At the same time the correction
due to large-scale deviations of boundary surfaces from the perfect shape  
$\Delta_{l}F_C^{(0)}$ may achieve
$\pm 0.3F_C^{(0)}\approx \mp 9.3\times 10^{-11}\,$N.
All these corrections are of the same order or much larger than the absolute error of
force measurements $\Delta F$. Moreover, the largest correction due to
large-scale deviations can not be estimated theoretically because the
actual shape of interacting bodies was not investigated in experiment [20].
In this situation the cancellation of contributions from different
corrections to the force value occurs very likely.

The constraints on the parameters of hypothetical interactions $\alpha$,
$\lambda$, $\lambda_n$ may be calculated from the inequality
\be
\vert F_C^{th}(a)+F_h(a)-F_C^{(0)}(a)\vert\leq\Delta F,
\label{46}
\ee
\noindent
where $F_C^{th}$ is the theoretical Casimir force value with account of all
corrections, $F_h$ is the hypothetical force $F_{Yu}$ or $F_n$
calculated in Secs.~III,~IV.

Substituting the general expression for  $F_C^{th}$ into (46) one obtaines
\be
\vert F_h(a)+\Delta_{\Sigma}(a)+\Delta_lF_C^{(0)}(a)\vert\leq\Delta F,
\label{47}
\ee
\newpage
\noindent
where
\be
\Delta_{\Sigma}(a)\equiv \Delta_{T}F_C^{(0)}(a)+\Delta_{\delta_0}F_C^{(0)}(a)+
\Delta_{d}F_C^{(0)}(a)
 \label{48}
\ee
\noindent
is the sum of corrections whose values for different values of $a$ are
accessible.

Although the values of the correction due to large-scale
 deviations are unknown its
dependence on $a$ is given by Eq.~(12). Thus for two different values of
$a$ it holds
\be
\Delta_{l}F_C^{(0)}(a_2)=\frac{1}{k_{21}}\Delta_{l}F_C^{(0)}(a_1),
\qquad k_{21}=\left(\frac{a_2}{a_1}\right)^4.
 \label{49}
\ee

According to the results of Sec.~III the value of the hypothetical force is
proportional to interaction constant $F_h(a_i)\equiv \alpha G_{\lambda}(a_i)$,
or to $\lambda_n$ for power-law interactions (see Sec.~IV), with some
known functions $G_{\lambda}$ ($G_n$). Considering (47) for two different
values of distance $a_1,a_2$ with account of (49)
and excluding the unknown quantity $\Delta_{l}F_C^{(0)}(a_1)$ we obtain the 
desired constraints for $\alpha$ ($\lambda_n$)
\bea
&&-(k_{21}+1)\Delta F-\Delta_{\Sigma}(a_1)+k_{21}\Delta_{\Sigma}(a_2)
\nonumber \\
&&\phantom{aaaaa}\leq
\alpha\left[G_{\lambda}(a_1)-k_{21}G_{\lambda}(a_2)\right]
 \label{50}  \\
&&\phantom{aaaaaaaaaa}\leq
(k_{21}+1)\Delta F-\Delta_{\Sigma}(a_1)+k_{21}\Delta_{\Sigma}(a_2).
 \nonumber
\eea

The specific values of $a_1,a_2$ in (50) should be chosen in the interval
$1\,\mu$m$\,\leq a\leq 6\,\mu$m in order to obtain the strongest constraints
on $\alpha, \lambda_n$. For the upper limit of the distance interval
($a\approx 6\,\mu$m) 
the Casimir force $F_C(a)$ from (14), i.e. together with the temperature
correction, should be considered as a
force under measuring. All corrections
to it are much smaller than $\Delta F=10^{-11}\,$N. By this reason for
such values $a$ the constraints on the hypothetical interaction may be 
obtained,
instead of (47), 
from a simplified inequality:
\be
\vert F_h(a)\vert =\vert \alpha G_{\lambda}(a)\vert\leq\Delta F.
 \label{51}
\ee

Now let us turn to numerical calculations of constraints starting with the
Yukawa-type potential. The constraints on $\alpha$ were obtained 
from Eq.~(50). For every $\lambda$ some pair of distances $a_1,a_2$ was
selected which provides us with the strongest constraints. The hypothetical
force $\alpha G_{\lambda}(a_i)$ was calculated by Eqs.~(29) for small 
$\lambda$, (32) for large $\lambda$ and numerically in the intermediate 
region according to
Sec.~III. The results are presented in Fig.~3 by the solid curve~1 which
corresponds to attractive hypothetical force ($\alpha>0$) and by the dotted
curve~2 corresponding to repulsion ($\alpha<0$). In the
($\lambda,\alpha$)-plane the region above the curves is prohibited and 
the region below the curves is permitted. Let us consider firstly the case 
$\alpha>0$.  Here the constraints for  $\lambda\sim 10^{-7}$m
were obtained by using $a_1= 1\,\mu$m, $a_2= 1.5\,\mu$m. With increase of
$\lambda$ the best constraints follow from $a_1= 1\,\mu$m, $a_2= 2\,\mu$m.
For $\lambda > 8\times 10^{-6}$m the values $a_1= 1\,\mu$m, $a_2= 3\,\mu$m
were used. Now let us turn to the case $\alpha<0$.
Considering $\lambda\approx 10^{-7}$m, we have chosen 
$a_1= 1\,\mu$m, $a_2= 1.5\,\mu$m once more. Then in the range till
$\lambda\approx 10^{-5}\,$m the values  $a_2= 2\,\mu$m or $a_2= 1.5\,\mu$m
were used ($a_1= 1\,\mu$m). For $\lambda > 10^{-5}\,$m the system of
inequalities (50) does not lead to better results than the single
inequality (51) used with $a= 6\,\mu$m.  
The complicated character of curves~1,2 in Fig.~3 (non-monotonic behaviour
of their first derivatives) is explained by the flaky structure of test bodies.
For $\lambda > 10^{-5}\,$m the metallic layers do not contribute essentially to the value of the force
and it is determined mostly by quartz. But for
$\lambda \lsim 10^{-5}\,$m the contribution of the layers becomes the main one.

In Fig.~3 the known constraints are shown also following from the former
Casimir force measurements between dielectrics [4--7] (curve~3),
Cavendish-type experiments (curve~4 [16], curve~5 [14], curve~6,
$\alpha >0$ [14],  curve~7,
$\alpha <0$ [15]) and E\"{o}tv\"{o}s-type experiment [17] (curve~8). It is
seen that the new constraints following from [20] are the best ones within a wide
range $2.2\times 10^{-7}$m$\,\leq\lambda\leq1.6\times 10^{-4}$m. They
surpass the old ones up to a factor of 30. For 
$\lambda <2.2\times 10^{-7}$m the old Casimir force measurements lead to 
better constraints than the new one. This is caused by the smallness of the
Casimir force between dielectrics comparing the case of metals and also
by the fact that there the force was measured for smaller values of $a$.

Now let us obtain constraints for the constants $\lambda_n$ of power-law 
hypothetical interactions which follow from the experiment [20]. Here the
inequalities (50) should be used once more, where $\alpha$ has to be
replaced by $\lambda_n$ and $G_{\lambda}(a_i)$ by $G_n(a_i)$.
For the power-law interactions the hypothetical force
$\lambda_nG_n(a_i)$ is calculated by Eqs.~(37),~(44),~(45) of Sec.~IV. The
strongest constraints are obtained for $a_1 =1\,\mu$m, $a_2= 3\,\mu$m.
Thus for the potential (33) with $n=2,~3,~4$ the new constraints are:    
$\lambda_2\leq 1.1\times 10^{-26}$, $\lambda_3\leq 1.6\times 10^{-14}$,
$\lambda_4\leq 3.6\times 10^{-3}$. Note that these constraints are not so
strong as the ones obtained from the old Casimir force measurements
($\lambda_2\leq 5\times 10^{-28}$, $\lambda_3\leq 5\times 10^{-15}$,
$\lambda_4\leq 3\times 10^{-3}$ [9]) or as the best ones  obtained up to day 
from the Cavendish-type experiment
 ($\lambda_2\leq 7\times 10^{-30}$, $\lambda_3\leq 7\times 10^{-17}$,
$\lambda_4\leq 1\times 10^{-3}$ [19]). The reasons for this are the same as 
for Yukawa-type interactions with small $\lambda$. Actually, the power-law
potential with $n=2$ leads to the force (45) between a lens and a disc
which does not depend on $a$. The dependence of $F_3$ on $a$ is also very week
(see (37)). By this reason the covering metallic layers do not contribute to 
the value of the hypothetical force and can not compensate the factors 
mentioned above. For $n=4$ there is the noticeable dependence on $a$ in (44).
By this reason the constraints obtained from the new experiment is almost 
of the 
same strength as from the old one.

As it is shown in the next Section the new constraints obtained from the
experiment [20] can be considerably strengthened owing to some
modification of its parameters.  

\newpage
\section{POSSIBLE IMPROVEMENT OF THE \protect\\OBTAINED RESULTS}

The experiment \cite{20} was designed to demonstrate the Casimir force between
metallic surfaces. The strengthening of the constraints on the constants of
hypothetical interactions was derived from its results
afterwards. Therefore it is likely that modifications of the design would 
allow to get much stronger
constraints. The simplest suggestion follows from the fact
that the hypothetical forces are proportional to the product of the densities
of the lens and the disc (see Secs.~III,~IV). The value of the Casimir force
measured in \cite{20} was determined by the thin covering gold layers only. For
the hypothetical forces which decrease with distance not so fast this is not
the case. By this reason the contribution of the hypothetical forces may be
increased by the use of some high density metals as material for the lens
and the disc instead of quartz. As such material iridium 
($\tilde{\rho}=22.4\,
$g/cm$^{3}$) looks very promising. The gold cover of $0.5\mum$ thickness
should be preserved due to its good conductivity. With a lens and a disc made
of iridium the values of the hypothetical forces (if any) are increased
approximately by a factor of $10^2$. An increase of the lens and the disc geometrical
parameters ($R$, $H$, $L$, $D$) to become larger than the values used in the
experiment \cite{20} is not required. According to our estimates this would
not lead to an essential further strengthening of the constraints (because the
additional volume is situated too far from the nearest points of the bodies).

In Fig. 3 (curve~9 for $\alpha<0$, curve~10 for $\alpha>0$) 
the constraints for Yukawa-type interactions are shown
which would follow from an experiment like \cite{20} with the lens and the
disc made  out of iridium covered by a gold layer. The constraints were
obtained using the inequality \Ref{50} in the same way as in Sec.~V for quartz
test bodies. The hypothetical force was calculated according to Sec. III with
$\rho_1=\rho'_{1}=\tilde{\rho}$. It is seen that the prospective  
constraints of curves~9,10 
are about 100 times stronger than that obtained actually (curves~1,2) in the
range $\lambda> 10^{-3}\m$. For smaller $\lambda$ the strengthening is
not so high because the gold  covering layers contributed hypothetical forces
essentially in this range in \cite{20}. For $\lambda \lsim 10^{-6}\m$ the
prospective constraints are approximately the same as the actually obtained 
from \cite{20} (here the gold layers themselves determine the result).
It is noticeable that with iridium test bodies the Casimir effect would give
the best  constraints for the wider $\lambda$-range
$2.2\times 10^{-7}\m\le\lambda\le 3.2\times 10^{-3}\m$ and would exceed 
the constraints
following from the Cavendish-type experiment of Ref. \cite{16} (curve 4).

Better constraints may also be obtained on the constants of power-law
hypothetical forces by use of iridium test bodies. Performing the calculations
using the inequalities \Ref{50} with $a_{1}=1\mum$, $a_{2}=3\mum$ and
Eqs. \Ref{37}, \Ref{45} for the hypothetical forces one obtains
$\lambda_{2}\le 1.2\times 10^{-28}$, $\lambda_{3}\le 1.8\times 10^{-16}$. In
the same way for $n=4$ with  $a_{1}=1\mum$, $a_{2}=2.5\mum$
one has 
$\lambda_{4}\le 8.4\times 10^{-5}$. All these constraints are stronger than 
those
obtained before from the Casimir force measurements between dielectrics (see
Sec. V). For $n=4$ the prospective constraint  is stronger than the best one 
obtained up 
today from the Cavendish-type experiment ($\lambda_{4}\le 1\times
10^{-3}$).

Now let us consider one more modification of the experiment \cite{20} which
causes a further strengthening of the constraints. The experiment \cite{20}
was not so sensitive as it might be. This was because of the missing vibration
isolation from the surrounding building. Using an appropriate isolation and
larger distances between the test bodies the
absolute sensitivity might reach the $0.01$ microdyne level \cite{37}. 

Here we estimate the prospective constraints which might follow from the
Casimir force measurements between the iridium lens and disc covered by gold
layers, assuming $\Delta F =10^{-13}\,$N. We consider the upper limit of the
$a$-interval, $a=6\mum$, and suggest that the theoretical value of the Casimir
force \Ref{14} including the temperature part is confirmed with the absolute
error $\Delta F$.  (Note that for $a=6\mum$ we have 
$F_C^{(0)}\approx -1.43\times
10^{-12}\,$N and $\Delta_T F_C^{(0)}\approx -2.48\times
10^{-12}\,$N, 
so that the temperature part can not be considered as a 
correction any more.) The modulus of
corrections to the Casimir force due to the finite conductivity and surface
distortions are less than $\Delta F$
 at $a =6 \mum$ (the largest of them would be
$\Delta_{\delta_{0}}F_{c}\approx0.36\times10^{-13}\,$N, see Fig.~2). 
That is why, as the first approximation, it is possible to get
the prospective constraints on the hypothetical force from the 
inequality \Ref{51}.

The calculational results are shown in Fig. 3 by the curve 11. It is seen that
the new prospective constraints overcome almost all results following from the
different Cavendish-type experiments (except of a small part of the curve 7
for $\alpha < 0$). They may become the best ones in a wide $\lambda$-range
$5\times10^{-7}\m\le\lambda\le5\m$.
The maximal strengthening comparing the results following from \cite{20} 
achieves
$10^4$ times. For the intermediate $\lambda$-range the strengthening achieves
several thousand times. For extremely small $\lambda$ the promised results
are weeker; this
is caused by the use of inequality \Ref{51} instead of the more exact one,
Eq. \Ref{50}. Strictly speaking, the
special investigation of large-scale deviations of
the surface from the ideal shape 
and short-scale distortions is needed to obtain the most strong constraints
on the constants of hypothetical interactions from the Casimir force
measurements. 

The prospective constraints of curve~11 would give the possibility to restrict
the masses of the spin-one antigraviton (graviphoton) and dilaton. The
interaction constants of Yukawa-type interaction due to exchange by these
particles are predicted by the theory: $\alpha_{a}\approx 10^{-40}$,
$\alpha_{d}\approx 2\times10^{-39}$ \cite{12}. 
As it is seen from curve 11 of Fig. 3, e.g., for graviphotons the permitted
values of $\lambda$ are $\lambda_{a}\le4\times10^{-4}\m$ or for its mass
$m_{a}\ge\hbar /(\lambda_{a}^{\max}c)\approx 5\times10^{-4}\,$eV.
These constraints are stronger than those known up to date
($\lambda_{a}<3\times10^{-3}\m$, $m_{a}\ge6\times10^{-5}\,$eV \cite{12})
obtained from Cavendish-type experiment. Note that obtaining much
stronger constraints for the parameters of the graviphoton is of special 
interest in connection with the recently claimed 
experimental evidence for the existence of this
particle from geophysical data \cite{38}.

Decreasing of the absolute error of force measurements till $\Delta F =
10^{-13}\,$N will give the possibility to strengthen the constraints on
power-law interactions as well. For a lens and a disc made of iridium and
covered by gold layers the results are obtained from the inequality \Ref{51}
with $a=6\mum$: $\lambda_{2}\le2.13\times10^{-30}$, 
$\lambda_{3}\le3.25\times10^{-18}$, $\lambda_{4}\le1.8\times10^{-6}$.
These constraints overcome essentially the current results following from the
Cavendish-type experiment (see Sec. V). The greatest strengthening by
several hundred times takes place for $n = 4$. 
That is why the realization of the
experiment on demonstration of the Casimir force with improved parameters is
of great interest for the problem of hypothetical interactions.


\section{ CONCLUSION AND DISCUSSION}

In this paper we performed a careful calculation of the Casimir and
hypothetical forces according to the configuration of experiment [20], 
i.e. between a
spherical lens and a disc made of quartz whose surfaces were covered by thin
layers of copper and gold. The finiteness of the disc area was taken into
account and the corrections were calculated to the Casimir force between a
lens and a disc of the infinite area (Sec.~II). These corrections were 
shown to be
negligible. That is why the use of the theoretical result for infinite disc in
[20] for confronting with experimental data is justified.

Different corrections to the Casimir force were discussed, e.g., due to the
finite conductivity of the boundary surfaces, deviations of their geometry
from the perfect shape and due to non-zero temperature (Sec.~II). It was shown
that the corrections due to finite conductivity and to short-scale distortions
have the opposite sign and may partly compensate each other. At the same time
the global deviations of the boundary surface geometry from the perfect shape
lead to both positive and negative corrections (which may reach $30\%$ of
the Casimir force acting in a perfect configuration  with space separation 
$a = 1\mum$). By this reason a detailed investigation of the boundary
surface geometry is required when confronting experimental and theoretical
results for configurations with a small space separation. As to the temperature
contribution to the Casimir force it may be considered as a correction for the
small space separation only and should be included into the force under
measuring starting from $a \approx 3\mum$.

The calculation of the Yukawa-type hypothetical force has shown that for 
large values of
$\lambda \rsim 10^{-3}\m$ it practically does not depend on $a$, for $\lambda
\rsim 10^{-5}\m$ the covering metallic layers do not contribute to its value
essentially, but for smaller $\lambda$ the contribution of the layers becomes
the main one (Sec. III). For the power-law interactions the strongest
dependence of the force on $a$ was obtained for $n = 4$ (Sec.~IV). 
For $n = 2$ it
is practically absent and for $n = 3$ there is only a weak dependence of the
force on $a$. As a result the covering metallic layers practically do not
contribute the value of force for $n = 2,3$.

The careful calculation of the Casimir and hypothetical forces in the
configuration of experiment [20] gave the possibility to obtain  reliable
constraints on the parameters of hypothetical interactions (Sec.~V). In the
case of Yukawa-type hypothetical interactions the new constraints surpass the
old ones following from the Casimir force measurements between dielectrics 
by a factor of 30 in a wide range 
$2.2\times10^{-7}\m\le\lambda\le1.6\times 10^{-4}\m$
(curves 1,2 in Fig.~3). In this $\lambda$-range the obtained constraints are
the best ones on the Yukawa-type interactions following from
laboratory experiments. For the power-law interactions the experiment [20]
does not lead to new constraints which would be better than the ones known up
to date.

According to the analysis presented above, by some modification of parameters
of the
experiment  [20] the related constraints on the constants of hypothetical
interactions may be essentially improved (Sec.~VI). With the use of iridium
test bodies (instead of quartz ones) the possible improvement is shown by the
curves 9,~10 in Fig. 3 and achieves two orders of magnitude. If in addition the
accuracy will be improved due to the use of vibrational isolation from the
surrounding building the resulting constraints are given by the curve 11 in
Fig.3. In this case the improvement will achieve four orders of magnitude for
large $\lambda$ and several thousand times for intermediate $\lambda$. It is
notable that the improved Casimir force measurement promises more strong
constraints for the range $1.6\times10^{-4}\m\le\lambda\le5\m$
than the ones known up to date from the Cavendish-type experiments. 
Obtaining
such constraints would supply us with new information about light
hypothetical particles, e.g., graviphoton, dilaton, scalar axion etc. The
experiment with iridium test bodies and improved accuracy will give the
possibility to strengthen constraints on power-law interactions as
well. Comparing the current constraints following from the Cavendish-type
experiment the prospective strengthening is by a factor of  3.3, 21.5 and 
555  for $n =2, 3, 4$, respectively.

To conclude, we would like to emphasize that the new measurements of the
Casimir force are interesting not only as the confirmation of one of the most
interesting predictions of quantum field theory (there are also other
experiments on Casimir effect in preparation, see, e.g., [39]). The additional
interest arises from the hope that it may well become a new method for the
search of hypothetical forces and associated light and massless 
elementary particles.

\section*{Acknowledgments}

The authors are greatly indebted to S.~K. Lamoreaux for additional information
about his experiment and several helpful discussions, especially concerning
the accuracy of force measurements. They also thank E.R. Bezerra de Mello,
V.B. Bezerra, G.T. Gillies and C. Romero for collaboration on the previous
stages of this investigation. G.L.K. and V.M.M. are indebted to the Institute
of Theoretical Physics of Leipzig University, where this work was
performed, for kind hospitability. G.L.K. was supported by  Saxonian
Ministry for Science and Fine Arts; V.M.M. was
supported by Graduate College on Quantum Field Theory at
Leipzig University.

\setcounter{equation}{0}
\renewcommand {\theequation}{A\arabic{equation}}
\section*{APPENDIX}

In this Appendix we present the derivation of several mathematical
expressions used in Secs.~II--IV. Let us start with the calculation of the
multiple integrals
\begin{eqnarray}
&&I_{2k+1}\equiv
\int\limits_{0}^{2\pi}\!d\varphi_1\!
\int\limits_{0}^{\theta_1}\!\sin\vartheta_1\, d\vartheta_1\!\!\!
\int\limits_{R_{\min}^{\prime}(\vartheta_1)}^{R_{\max}^{\prime}(\vartheta_1)}
\!\!\!r_1^2\, dr_1
\label{A1} \\
&&\phantom{aaaaaa}\times
\int\limits_{0}^{2\pi}\!d\varphi_2\!
\int\limits_{0}^{\theta_2}\!\sin\vartheta_2\, d\vartheta_2\!\!\!
\int\limits_{R_{\min}(\vartheta_2)}^{R_{\max}(\vartheta_2)}\!\!\!
r_2^2\, dr_2\,
r_{12}^{-2k-1},
\nonumber
\end{eqnarray}
\noindent
whose  integration limits satisfy 
the conditions
\begin{eqnarray}
&&R_{\min}>R_{\max}^{\prime}, \qquad
\sin\theta_1\ll 1, \qquad \sin\theta_2\ll 1,
\nonumber \\
&&R_{\max}-R_{\min}^{\prime}\ll R_{\min}^{\prime}.
\label{A2} 
\end{eqnarray}
\noindent
Integrals of that type were essential in Sec.~II (for $k=3$) and in
Sec.~IV (for $k=1$).

To calculate $I_{2k+1}$ it is convenient to use the following expansion
into a series of spherical harmonics [34]
\bea
&&r_{12}^{-2k-1}=4\pi
\sum\limits_{l=0}^{\infty}
\frac{1}{2l+1}\,
{a}_{l}^{[-(2k+1)]}(r_1,r_2)\,
\label{A3} \\
&&\phantom{aaaaaa}\times
\sum\limits_{m=-l}^{l}
{Y}_{lm}^{\ast}(\vartheta_1,\varphi_1)\,
{Y}_{lm}(\vartheta_2,\varphi_2),
\nonumber
\eea
\noindent
where the radial part may be represented in the form
\begin{eqnarray}
&&{a}_{l}^{[-(2k+1)]}(r_1,r_2)=
\frac{\left(\frac{2k+1}{2}\right)_l}{\left(\frac{1}{2}\right)_l}\,
\frac{(r_1\,r_2)^l}{(r_1+r_2)^{2l+2k+1}}
\label{A4} \\
&&\phantom{aaaaaa}\times
F\!\left(l+\frac{2k+1}{2},l+1;2l+2;\frac{4\,r_1\,r_2}{(r_1+r_2)^2}\right).
\nonumber
\end{eqnarray}
\noindent
Here $F(\alpha,\beta;\gamma;z)$ is the hypergeometric function,
$(n)_l\equiv\Gamma(l+n)/\Gamma(n)$, $\Gamma(z)$ is the gamma function, and
in accordance with inequalities (A2) $r_2>r_1$.

It is readily seen that due to (A2) the argument 
$z\equiv  4r_1r_2/(r_1+r_2)^2$ of the hypergeometric 
function in (A4)  is of order of unity. So
according to [35] it is profitable to use the hypergeometric function of
the argument $z_1=1-z$:
\begin{eqnarray}
&&F\!\left(l+\frac{2k+1}{2},l+1;2l+2;z\right)=\frac{\Gamma(2l+2)\,
\Gamma\left(\frac{1-2k}{2}\right)}{\Gamma\left(l+\frac{3-2k}{2}\right)\,
\Gamma(l+1)}
\nonumber \\
&&\phantom{aaaaaaa}\times
F\!\left(l+\frac{2k+1}{2},l+1;\frac{2k+1}{2};z_1\right)
\label{A5} \\
&&\phantom{aa}+
\left(\frac{r_2+r_1}{r_2-r_1}\right)^{2k-1}\frac{\Gamma(2l+2)\,
\Gamma\left(\frac{2k-1}{2}\right)}{\Gamma\left(l+\frac{2k+1}{2}\right)\,
\Gamma(l+1)}
\nonumber \\
&&\phantom{aaaaaaa}\times
F\!\left(l+\frac{3-2k}{2},l+1;\frac{3-2k}{2};z_1\right).
\nonumber  
\end{eqnarray}

Note that due to inequalities (A2) it holds $z_1\ll 1$. On account of this
the first contribution on the right-hand side of (A5) is of order
$z_1^{k-\frac{1}{2}}$ relatively the second one and it is possible to omit it
for $k\geq 1$. One can also substitute the hypergeometric function of the
small argument $z_1$ in the second contribution to (A5) by unity. In addition
the product of radiuses in (A4) may be expressed in terms of their sum with
the same accuracy
\begin{equation}
r_1\,r_2=\frac{1}{4}\left[(r_1+r_2)^2-(r_1-r_2)^2\right]
\approx \frac{1}{4}\,(r_1+r_2)^2. 
\label{A6}
\end{equation}

Substituting (A5) and (A6) into (A4) and using the properties of gamma
function we obtain
\begin{equation}
{a}_{l}^{[-(2k+1)]}(r_1,r_2)\approx
\frac{2\,(2l+1)}{(2k-1)\,(r_1+r_2)^2\,(r_2-r_1)^{2k-1}}.
\label{A7}
\end{equation}

As a result expansion (A3) with the condition $k\geq 1$ takes the form
\bea
&&r_{12}^{-2k-1}=\frac{8\pi}{2k-1}\,\frac{(r_2-r_1)^{1-2k}}{(r_1+r_2)^2}
\label{A8} \\
&&\phantom{aaaaaa}\times
\sum\limits_{l=0}^{\infty}
\sum\limits_{m=-l}^{l}
{Y}_{lm}^{\ast}(\vartheta_1,\varphi_1)\,
{Y}_{lm}(\vartheta_2,\varphi_2).
\nonumber
\eea

After the substitution of (A8) with the use of completeness relation
for the spherical harmonics [34]
\bea
&&\sum\limits_{l=0}^{\infty}
\sum\limits_{m=-l}^{l}
{Y}_{lm}^{\ast}(\vartheta_1,\varphi_1)\,
{Y}_{lm}(\vartheta_2,\varphi_2)=\delta (\varphi_1-\varphi_2)\,
\nonumber \\
&&\phantom{aaaaaa}\times
\delta (\cos\vartheta_1-\cos\vartheta_2)
\label{A9}
\eea
Eq.~(A1) may be represented as
\begin{eqnarray}
&&I_{2k+1}=\frac{8\pi}{2k-1}
\int\limits_{0}^{2\pi}\!d\varphi_1\!
\int\limits_{0}^{2\pi}\!d\varphi_2\,\delta (\varphi_1-\varphi_2)\!
\int\limits_{0}^{\theta_1}\!\sin\vartheta_1\, d\vartheta_1\!\!
\nonumber \\
&&\phantom{aa}\times
\int\limits_{0}^{\theta_2}\!\sin\vartheta_2\, d\vartheta_2
\delta (\cos\vartheta_1-\cos\vartheta_2)
\label{A10} \\
&&\phantom{aaaa}\times
\int\limits_{R_{\min}^{\prime}(\vartheta_1)}^{R_{\max}^{\prime}(\vartheta_1)}
\!\!dr_1\!
\int\limits_{R_{\min}(\vartheta_2)}^{R_{\max}(\vartheta_2)}\!\!
dr_2\,
\frac{r_1^2\,r_2^2}{(r_1+r_2)^2\,(r_2-r_1)^{2k-1}}.
\nonumber
\end{eqnarray}

After the integration with $\delta$-functions in (A10) the result is
\bea
&&I_{2k+1}=\frac{16\pi^2}{2k-1}
\int\limits_{0}^{\min(\theta_1,\theta_2)}\!\!\!\sin\vartheta\, d\vartheta
\label{A11}  \\
&&\phantom{aaaa}\times
\int\limits_{R_{\min}^{\prime}(\vartheta)}^{R_{\max}^{\prime}(\vartheta)}
\!\!dr_1\!
\int\limits_{R_{\min}(\vartheta)}^{R_{\max}(\vartheta)}\!\!
dr_2\,
\frac{r_1^2\,r_2^2}{(r_1+r_2)^2\,(r_2-r_1)^{2k-1}}.
\nonumber
\eea

Eq.~(A11) is useful for the calculation of the Casimir force (Sec.~II) and
power-law hypothetical interaction decreasing as the third power of
distance (Sec.~IV). 

Now let us calculate the integral (31) which expresses the asymptotic
behaviour of Yukawa-type interaction for large $\lambda$ (Sec.~III). We
use once more the expansion (A3) into the spherical harmonics with $k=1$.
For the radial part, instead of (A4), it is more convenient to apply
the equivalent representation [34]
\begin{equation}
{a}_{l}^{(-3)}(r_1,r_2)=\frac{2l+1}{r_2^2-r_1^2}
\frac{r_1^l}{r_2^{l+1}}=
\frac{2l+1}{r_2(r_2^2-r_1^2)}\left(\frac{r_1}{r_2}\right)^l.
\label{A12}
\end{equation}

To lowest order in the small parameters $H/R$, $D/R$ it holds
$r_1/r_2\approx\cos\vartheta_2$.
Note also that only the term with $m=0$ from (A3) gives non-zero
contribution when integrating with respect of $\varphi_{1,2}$ in (31).
As a result (31) may be rewritten as 
\begin{equation}
F_{Yu}\approx -\alpha\,\frac{\hbar c}{m_p^2}\,\rho_1^{\prime}\,\rho_1
\int\limits_{0}^{\theta_2}\sin\vartheta_2\, d\vartheta_2\,
\Phi(\cos\vartheta_2),
\label{A13}
\end{equation}
\noindent
where the following notation is introduced
\begin{eqnarray}
&&\Phi(\cos\vartheta_2)=4\pi^2
\int\limits_{0}^{\theta_1}\sin\vartheta_1\, d\vartheta_1
\int\limits_{\frac{R-H}{\cos\vartheta_1}}^{R}\!\!r_1^2\, dr_1
\nonumber \\
&&\phantom{aaaaaa}\times
\int\limits_{\frac{R+a}{\cos\vartheta_2}}^{\frac{R+a+D}{\cos\vartheta_2}}
\!\!r_2\, dr_2\,
\frac{r_2\cos\vartheta_2-r_1\cos\vartheta_1}{r_2^2-r_1^2}
\nonumber \\
&&\phantom{aaaaaa}\times
\sum\limits_{l=0}^{\infty}
(\cos\vartheta_2)^l\,(2l+1)\,P_l(\cos\vartheta_1)\,P_l(\cos\vartheta_2),
\label{A14}
\end{eqnarray}
\noindent
and $P_l(z)$ are Legendre polynomials.

Now it is useful to change variables in (A13), (A14) according to
$t_1=\cos\vartheta_1$, $t_2=\cos\vartheta_2$, $x_2=t_2r_2$. After that
(A13), (A14) take the form
\begin{equation}
F_{Yu}\approx -\alpha\,\frac{\hbar c}{m_p^2}\,\rho_1^{\prime}\,\rho_1
\int\limits_{\cos\theta_2}^{1}\!dt_2\,
\Phi(t_2),
\label{A15}
\end{equation}
\noindent
where $\cos\theta_2=R/\sqrt{R^2+L^2}\approx 1-L^2/(2R^2)$   and
\begin{eqnarray}
&&\Phi(t_2)=4\pi^2
\int\limits_{\cos\theta_1}^{1}\!\! dt_1\!
\int\limits_{\frac{R-H}{t_1}}^{R}\!\!\!r_1^2\, dr_1\!\!\!
\int\limits_{R+a}^{R+a+D}
\!\!\!\!x_2\, dx_2\,
\frac{x_2-r_1\,t_1}{x_2^2-r_1^2\,t_2^2}
\nonumber \\
&&\phantom{aaaaaa}\times
\sum\limits_{l=0}^{\infty}
t_2^l\,(2l+1)\,P_l(t_1)\,P_l(t_2),
\label{A16}
\end{eqnarray}
\noindent
with $\cos\theta_1=(R-H)/R$.

It is not difficult to calculate the integral (A15) approximately taking into
account that $L^2/(2R^2)\ll 1$ and expanding $\Phi(t_2)$ into a Taylor series
near the point $t_2=1$:
\begin{eqnarray}
&&F_{Yu}\approx -\alpha\,\frac{\hbar \,c}{m_p^2}\,\rho_1^{\prime}\,\rho_1\,
\frac{1}{2}[\Phi(1)+\Phi(\cos\theta_2)]\frac{L^2}{2R^2}
\nonumber \\
&&\phantom{aaa}
\approx -\alpha\,\frac{\hbar c}{m_p^2}\,\rho_1^{\prime}\,\rho_1\,
\left[\Phi(1)-\frac{1}{2}\Phi^{\prime}(1)\frac{L^2}{2R^2}\right]
\frac{L^2}{2R^2},
\label{A17}
\end{eqnarray}
\noindent
where $\Phi^{\prime}(1)=(d\Phi(t_2)/dt_2)\vert_{t_2=1}$.

The value $\Phi(1)$ can be calculated with account of equality
\begin{equation}
\sum\limits_{l=0}^{\infty}
(2l+1)\,P_l(t_1)\,P_l(t_2)=2\,\delta(t_1-t_2),
\label{A18}
\end{equation}
\noindent
which follows from the completeness relation (A9). The result for the lowest 
order in small parameters $H/R$, $D/R$ is
\begin{equation}
\Phi(1)=4\,\pi^2\,R^2\,D\,H.
\label{A19}
\end{equation}

Now let us find the contribution of the second term from the right-hand side
of (A17). For this purpose we differentiate (A16)
\begin{eqnarray}
&&\Phi^{\prime}(1)=4\pi^2\!\!
\int\limits_{\cos\theta_1}^{1}\!\! dt_1\!
\int\limits_{\frac{R-H}{t_1}}^{R}\!\!\!r_1^2\, dr_1\!\!\!
\int\limits_{R+a}^{R+a+D}
\!\!\!\!x_2\, dx_2\,
\label{A20} \\
&&\phantom{a}\times
\left[
\vphantom{\left(\sum\limits_{l=0}^{\infty}\right)_{t-2=1}}
\frac{2(x_2-r_1\,t_1)r_1^2}{(x_2^2-r_1^2)^2}
\sum\limits_{l=0}^{\infty}
(2l+1)\,P_l(t_1)\,P_l(1)\right.
\nonumber \\
&&\phantom{aa}+\left.
\frac{x_2-r_1\,t_1}{x_2^2-r_1^2}\left(
\frac{d}{dt_2}
\sum\limits_{l=0}^{\infty}
t_2^l\,(2l+1)\,P_l(t_1)\,P_l(t_2)\right)_{t_2=1}\right].
\nonumber
\end{eqnarray}
The first contribution on the right-hand side of (A20) can be calculated
by the use of the completeness relation (A18) and for the lowest order in
$H/R$ and $D/R$ this results in
\begin{equation}
\Phi^{\prime}(1)\approx 4\pi^2\,R^3\,D\,
\left(\ln\frac{D+H}{D}+\frac{H}{D}\ln\frac{D+H}{H}\right).
\label{A21}
\end{equation}
It is easily seen that in the lowest order in $H/R$ and $D/R$ 
the second contribution to (A20) is proportional to $2\pi^2 H^2\,D\,R$, i.e.
is a quantity of the order $H^2/R^2$ comparing the first one (A21). By this
reason the second contribution on the right-hand side of (A20) may be 
neglected.

Substituting (A19) and (A21) into (A17) we come to the result (32) for the
asymptotic behaviour of the Yukawa-type hypothetical interaction in the
limit of large value of $\lambda$. 




\onecolumn

\begin{figure}[h]\psfig{figure=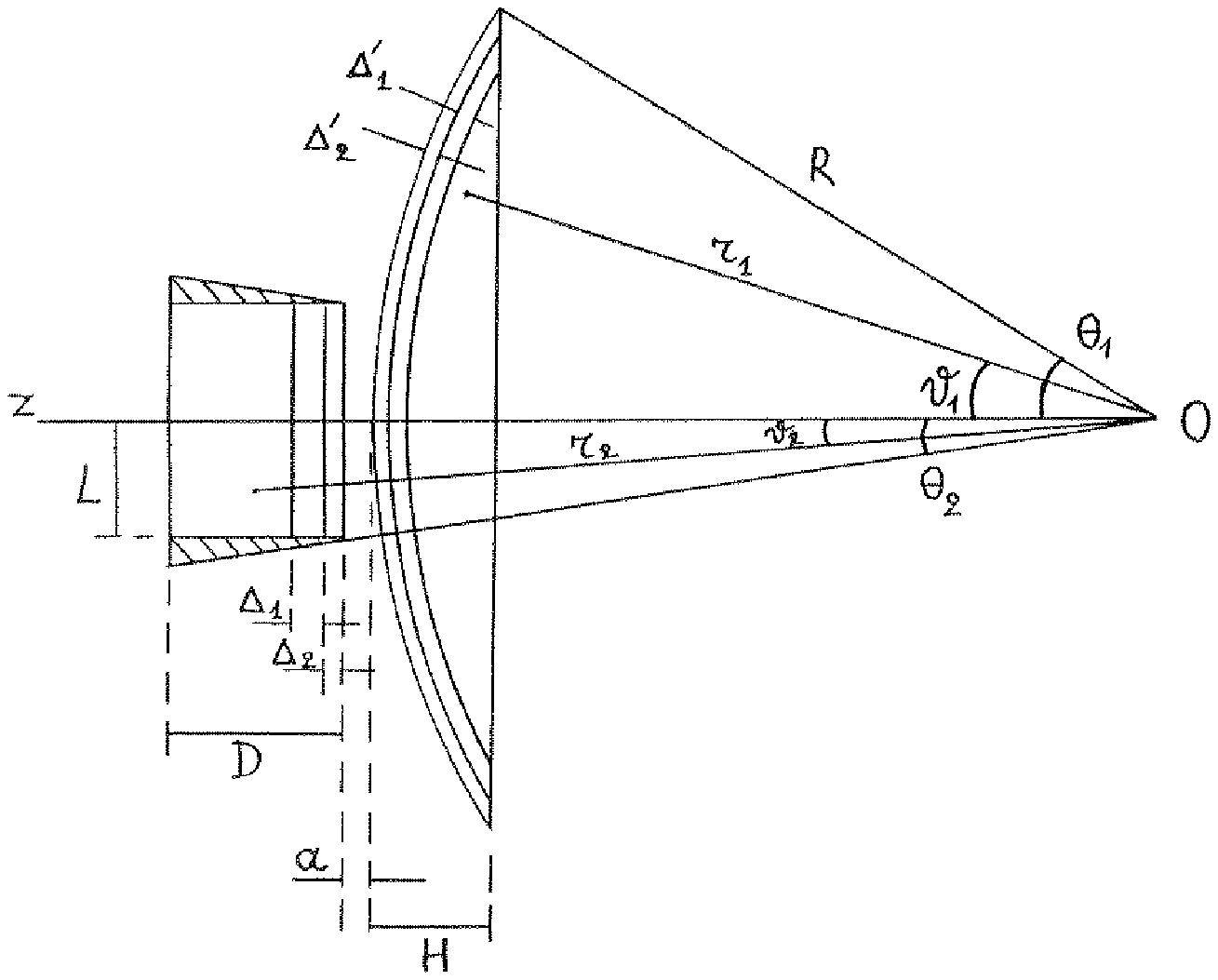}\\[12pt]
\caption{ Configuration of a spherical lens and a disc.
Point $O$ is the center of curvature of the lens with height $H$ and 
curvature
radius $R$. The thickness of the disc is $D$ and its radius is $L$,
$a$ being the distance between the disc and the lens. The thicknesses of
$Cu$ and $Au$ layers on the lens are $\Delta_1^{\prime}, \Delta_2^{\prime}$
and on the disc --- $\Delta_1, \Delta_2$ respectively. The small volume
added to the disc in calculations is shown by hatching.}
\end{figure}

 \begin{figure}[h]\psfig{figure=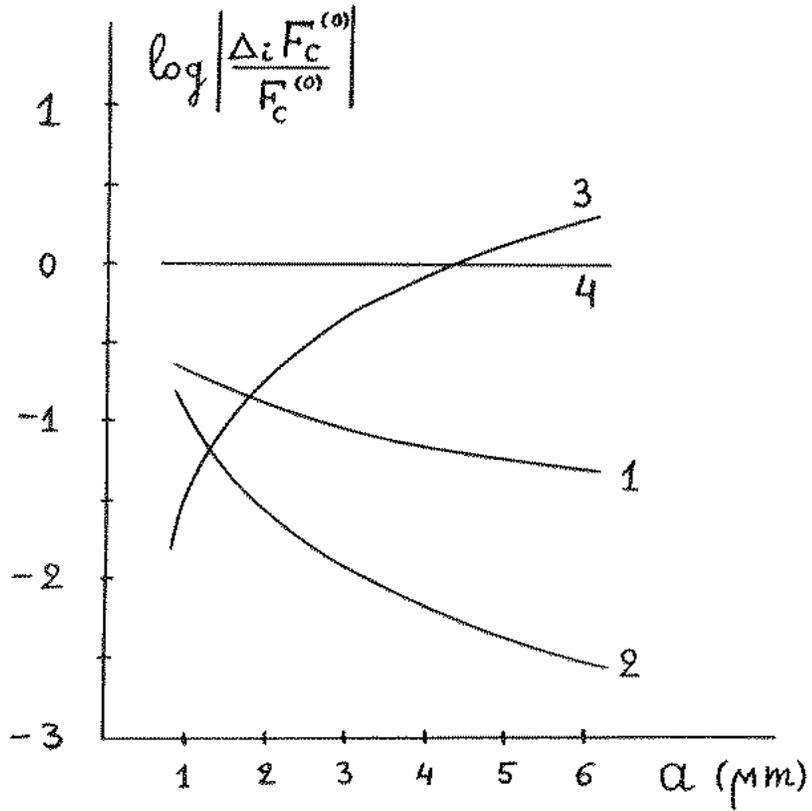}
\caption{The relative role of different corrections
to the Casimir force in configuration of a lens and a disc.
Curve~1  shows the correction due to the finite
conductivity ($\Delta_{\delta_0}F_C^{(0)}$), 
curve~2 shows the correction due to 
short-scale distortions ($\Delta_{d}F_C^{(0)}$), 
curve~3 shows the temperature correction ($\Delta_{T}F_C^{(0)}$). 
By the curve~4 the Casimir force itself is shown.}\end{figure}

 \begin{figure}[h]\psfig{figure=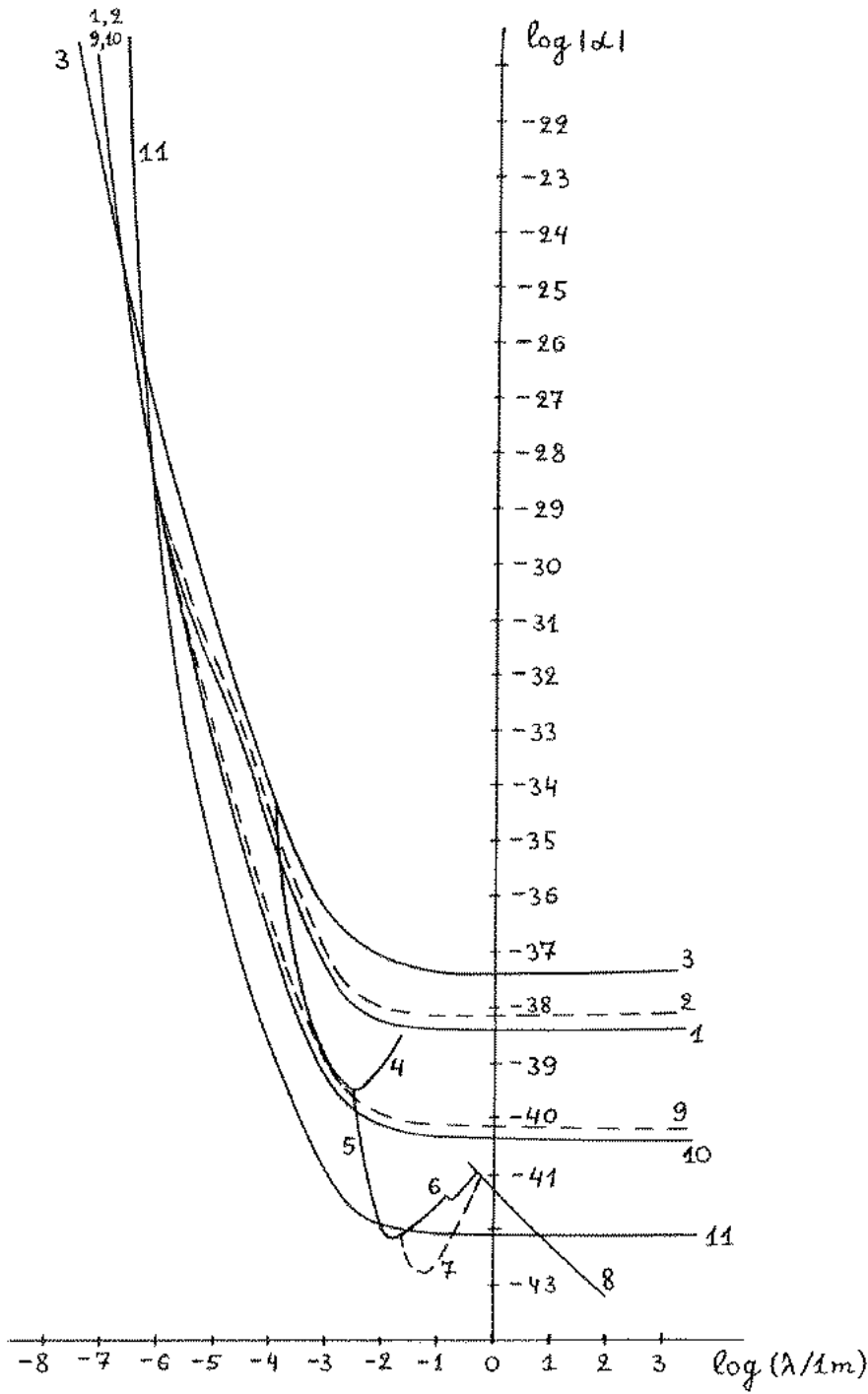}
\caption{Constraints for the constants of hypothetical
Yukawa-type interactions following from the force measurements.
Curve~1 follows from the new measurement of the Casimir force
($\alpha >0$),  curve~2 shows the same with $\alpha <0$, curve~3 results
from the old Casimir force measurements between dielectric bodies [4].
Constraints from Cavendish-type experiments are shown by 
curve~4 [16], curve~5 [14], curve~6, $\alpha >0$ [14],  curve~7,
$\alpha <0$ [15]. Curve~8 follows from the 
E\"{o}tv\"{o}s-type experiment [17]. Prospective constraints are shown by
the curves~9--11: from the Casimir force measurements with iridium test
bodies (curve~9, $\alpha >0$ and curve~10, $\alpha <0$), with iridium
test bodies and the improved accuracy (curve~11). The regions of
($\lambda,\alpha$)-plane below the curves are permitted and
above the curves are prohibited.  }\end{figure}

\end{document}